\documentclass[11pt]{article}
\usepackage{mathrsfs}
\usepackage{amssymb}
\usepackage{amsmath}
\usepackage{amsbsy}
\usepackage{epsfig}
\usepackage{enumerate}
\usepackage{bm}
\newtheorem{theorem}{Theorem}[section]

\newtheorem{assumption}{Assumption}[section]

\newtheorem{corollary}{Corollary}[section]

\newbox\TempBox \newbox\TempBoxA

\def\pr{\textsf{P}} 
\def\ep{\textsf{E}} 

\def\Cov{\textsf{Cov}} 
\def\Var{\textsf{Var}} 

\def\underwiggle 1{
\ifmmode\setbox\TempBox=\hbox{$ 1$}\else\setbox\TempBox=\hbox{
1}\fi \setbox\TempBoxA=\hbox to \wd\TempBox{\hss\char'176\hss}
\rlap{\copy\TempBox}\smash{\lower9pt\hbox{\copy\TempBoxA}} }
\renewcommand{\baselinestretch}{1.7}

\begin{document}

\thispagestyle{empty}

\begin{center}
 { \LARGE\bf  Response-adaptive randomization: an overview of designs and asymptotic theory}
\end{center}

\begin{center} {\sc
Li-Xin Zhang\footnote{Research supported by Grants from the National Natural Science Foundation of China (No. 11225104) and the Fundamental
Research Funds for the Central Universities.
}
}\\
{\sl \small Department of Mathematics, Zhejiang University, Hangzhou 310027 \\
and Zhejiang University City College} \\
Email:stazlx@zju.edu.cn
\end{center}




\baselineskip 22pt

\renewcommand{\baselinestretch}{1.7}



\section{ Introduction}
\setcounter{equation}{0}
In  clinical trials, patients normally arrive sequentially. Response-adaptive treatment allocation procedures are sequentially adaptive schemes that uses past
treatment assignments and patients' responses to select   future
treatment assignments. Historically, response-adaptive treatment allocation procedures were
developed for the purpose of assigning more patients to the empirically better treatment.
  Early important work on response-adaptive designs can be traced back to Thompson (1933) and Robbins (1952).
Since then, a lot of response-adaptive designs have been proposed in literature. A history of the subject is
discussed in Rosenberger and Lachin (2002) and Hu and Rosenberger (2006).  The most famous non-randomized
response-adaptive treatment allocation procedure is the play-the-winner (PW) rule proposed by Zelen (1969), in which a success on a treatment results in the next patient's assignment
to the same treatment, and a failure on the treatment results in the next patient's
assignment to the opposite treatment. The idea of incorporating randomization
into response-adaptive treatment assignments is due to Wei and Durham (1978),
who proposed a randomized play-the-winner (RPW) rule for binary
outcome trials. With this approach, each
patient's treatment assignment is determined by drawing a ball from an urn,
and the urn composition is updated by adding an additional ball of the same
type  if the patient's response is a success and an additional ball of the opposite type
 if the patient's response is a failure, so that the
ball corresponding to the better treatment has a larger probability to be drawn.
The RPW rule was used to design a pediatric trial of extracorporeal membrane
oxygenation (ECMO; Bartlett et al., 1985), which compared the ECMO therapy
versus the conventional therapy. Unfortunately, the trial provided very little
information about survival rates of the two treatments. The trial stopped after enrolling 12 infants, of whom one
infant was randomized to the conventional therapy and died and 11 infants
were randomized to ECMO treatment and all survived.
This design and subsequent
data analysis generated a lot of controversy and much of the criticism
of response-adaptive randomization (RAR) designs; in fact the ECMO trial example is still often used as a reason
not to perform response-adaptive randomization at all.
 According to the modern theory on response-adaptive randomization designs, the failure of the ECMO trial can be explained mainly
by  the trial's small sample size and the poor operating
characteristics of the RPW rule, in particular, the rule's high variability and
dependence on the initial composition of the balls in the urn.
Since the time of the ECMO trial, the body of knowledge on response-adaptive
randomization has grown significantly and many new methods have been
proposed to address past criticisms and concerns that have hindered the use of
adaptive randomization in clinical trials.
In this paper we provide an overview of important research works on response-adaptive
randomization completed in the past decades. In the next section, we give the general framework.
The RPW rule is only one example of a broad class
of randomized urn models and is generalized to the ``generalized P\'olya urn model'' (GPU).
Other
notable response-adaptive randomization urn designs with desirable statistical properties are proposed in the  past decades including
  the ``drop-the-loser (DL)
rule'' by Ivanova (2003), ``sequential estimated-adjusted urn'' (SEU) by Zhang, Hu and Cheung (2006),
 the ``generalized drop-the-loser (GDL) rule'' by Zhang et al. (2007), and the ``immigrated urn (IMU) model'' by  Zhang et al. (2011).
We state the theory on urn models in Section \ref{sectUrn}. Modern research on response-adaptive
randomization has focused on the development of optimal response-adaptive
randomization procedures that maintain or increase power over traditional balanced
randomization designs and minimize expected treatment failures. The optimization problem and the efficiency of
the response-adaptive randomization designs are discussed in Section \ref{sectOpEffic},
 where the methods and theoretical principles are proposed  for defining a  design with desirable or most desirable statistical properties.
 The asymptotic properties of several modern designs such as ``doubly adaptive biased coin design'' (DBCD)  of Hu and Zhang (2004a), ``efficient response adaptive randomization designs''(ERADE) of Hu, Zhang and He (2009) etc are presented.
 In section~\ref{sectSelBias}, the selection bias and  the lack of randomness are discussed. The measures for evaluating  the selection bias and the degree of the lack of randomness are defined and the measure values of several randomization procedures are given. Survival and delayed responses are briefly discuss in Section~\ref{sectDelay}. A theorem is given for showing that the mild delay will not effect the theoretical  results of response-adaptive randomization designs. The last section gives a concluding remark.

\section{Framework}\label{sectFrame}
\setcounter{equation}{0}
We consider   $K$-treatment clinical
trials, $K\ge 2$. Suppose that the outcome of treatment $k$ follows a probability distribution $f_k(x|\bm\theta_k)$ indexed by a parameter $\bm \theta_k$.
The patients arrive at the clinical trial sequentially and will be allocated to one of the $K$ treatments with a certain probability one by one.
 After the first $m$ assignments, the $(m+1)$-th
patient will be assigned to treatment $k$ with a probability
 $$p_{m+1,k}, \quad k=1,\ldots, K. $$
The probabilities $p_{m+1,k}$s are usually functions of the allocation results and outcomes of the treatments of previous $m$ trials.
Let $N_{m,k}$ be the number of the patients assigned to the treatment
$k$ in the first $m$ stages, $k=1,\ldots, K$, and   $\bm N_m=(N_{m,1},\ldots, N_{m,K})$.
Two problems are always concerned in the adaptive randomization studies. One is the way   for defining the allocation probabilities $p_{m+1,k}$s so that the design will  achieve the desired purposes in clinical trials. The other is the theory  on the behaviors of  $N_{m,k}$s and related statistical inference.

When $K=2$, in the play-the-winner  rule of Zelen (1969), the  allocation probability $p_{m+1,k}$ is defined to be $1$ if the $m$-th patient is assigned to treatment $k$ and success, and $0$ otherwise, $k=1,2$.  And in the randomized play-the-winner   rule of  Wei and Durham (1978), the allocation probability $p_{m+1,k}$ is defined to be the current proportion of balls of type $k$ in the urn:
$$ p_{m+1,k}=\frac {Y_{m,k}}{Y_{m,1}+Y_{m,2}}, $$
where $Y_{m,k}$ is the number of balls of type $k$ and is defined successively by $Y_{m,k}=Y_{m-1,k}+1$ if the $(m-1)$-th patient is assigned to treatment $k$ and successes, or  the $(m-1)$-th patient is assigned to the other treatment and fails.  For both the play-the-winner rule and randomized play-the-winner rule, it is shown that
$$ \frac{N_{n,k}}{n}\to \frac{q_k}{q_1+q_2}\;\; \text{ in probability},  $$
where $q_k$ is the failure rate of the treatment $k$, $k=1,2$.

\section{ Urn Models}\label{sectUrn}
\setcounter{equation}{0}

In the latter part of the 20th century, most research on response-adaptive
randomization focused on ad hoc designs based on stochastic processes such
as urn models (see Rosenberger, 2002). Many of these were extensions of the
randomized play-the-winner methodology, and  they focused on putting more patients on the superior treatment.
One large family is the family of  the generalized P\'olya urn models.

\subsection{Generalized P\'olya Urn} Consider an urn containing balls of $K$ types. At the beginning, the urn contains
$\bm Y_0=(Y_{0,1}, \ldots, Y_{0,K})$ particles, where $Y_{0,k}>0$
denotes the number of particles of type $k$, $k=1,\ldots, K$.  After $m$ assignments,  the urn composition is denoted by the row vector $\bm
Y_m=(Y_{m,1}, \ldots, Y_{m,K})$. The $(m+1)$-th patient is randomized to treatments by drawing a ball from the urn with replacement. If the ball drawn is of type $k$, then  the patient is assigned to treatment $k$, i.e., the allocation probability is defined as
$$ p_{m+1,k}= \frac{Y_{m,k}}{|\bm Y_m|}, \;\; k=1,\ldots, K, $$
where $|\bm Y_m|=Y_{m,1}+\cdots+Y_{m,K}$.   We then wait for
observing a random variable $\bm \xi(m+1)$, the response of the
treatment $k$ at the $(m+1)$-th patient.   After that,  additional $D_{k,q}(m+1)\ge 0$ balls of type
$q$, $q=1,\ldots, K$, are added to the urn, where $D_{k,q}(m+1)$ is a
function of $\bm \xi(m+1)$ and also may be a function of urn
compositions, assignments and responses of previous stages. This
procedure is repeated through out $n$ stages.
 This
relation can be written as the following recursive formula:
\begin{equation}\label{eq1.1}
 \bm Y_m = \bm Y_{m-1}+\bm X_m \bm D_m, \end{equation}
 where $\bm D_m=\big(D_{k,q}(m)\big)_{k,q=1}^K $, and $\bm X_m$ is the result
of the $m$-th draw, distributed according to the urn composition at
the previous stage, i.e., if the $m$th draw is a type $k$ ball,
then the $k$th component of $\bm X_m$ is $1$ and other components
are $0$. The matrices $\bm D_m$'s are referred to  as the   adding
rules. The conditional expectations
$\bm H_m=\big(\ep[D_{k,q}(m)\big|\mathscr{F}_{m-1}]\big)_{k,q=1}^d $,
  given the history sigma field $\mathscr{F}_{m-1}$ generated by
the urn compositions  $\bm Y_1,\ldots, \bm Y_{m-1}$, the assignments
$\bm X_1,\ldots, \bm X_{m-1}$ and the responses $\bm \xi(1),\ldots,
\bm \xi(m-1)$ of all previous stages, $m=1,2,\ldots$, are referred to as the
generating matrices.
 When $\bm
D_m$, $m=1,2,\ldots, $ are independent and identically distributed,
the GFU model is usually  said to be  homogeneous. In such case
$\bm H_m=\bm H$ are identical and nonrandom, and usually the
adding rule $\bm D_m$ is merely a function of the $m$th patient's
observed outcome. In the general heterogeneous cases, both $\bm
D_m$ and $\bm H_m$ depend on the entire history of all previous
trials which provides more information of the efficacy of the
treatments. However, we should suppose that $\bm H_m$ will converge to a matrix $\bm H$.

   The randomized play-the-winner rule is a homogeneous urn model with
$$ \bm D_m=\begin{pmatrix} \xi_{m,1} & 1-\xi_{m,1} \\1-\xi_{m,2} & \xi_{m,2}\end{pmatrix},\;\;
\bm H= \begin{pmatrix} p_1 & q_1 \\q_2 & p_2\end{pmatrix}, $$
where $\xi_{m,k}=1$ if the response of the $m$-th patient on   treatment $k$ is a success, and $=0$ for otherwise, $p_k$ is the success probability of treatment $k$, $q_k=1-p_k$, $k=1,2$.

For considering the asymptotic properties of $\bm N_n$, we suppose that   $\bm H$ has  a simple largest eigenvalue $\beta>0$  and
the corresponding  right eigenvector  $\bm v=(v_1,\ldots,v_K)$ and the left
eigenvector $\bm u^{\prime}=(u_1,\ldots, u_K)^{\prime}$ with $\sum_k
v_k=\sum_k v_k u_k=1$ and $ v_k>0, u_k>0$, $k=1,\ldots, K$.
Define $\Delta\bm M_{m,1}=\bm X_m-\ep[\bm X_m|\mathscr{F}_{m-1}]$,
$\Delta \bm M_{m,2}=\bm X_m(\bm D_m-\bm H_m)$ and $\overline{\bm H}=\bm H-\beta\bm 1^{\prime}\bm v$, where $\bm 1=(1,\ldots,1)$. It can be shown that $|\bm Y_m|\sim \beta m$ and
\begin{align}\label{eq1}
\Delta \bm Y_m=& \bm Y_m-\bm Y_{m-1}\\
=&\bm X_m(\bm D_m-\bm H_m)+(\bm X_m-\ep[\bm X_m|\mathscr{F}_{m-1}])\bm H_m+\frac{\bm Y_{m-1}}{|\bm Y_{m-1}|}\bm H_m\nonumber\\
=& \bm v+ \Delta \bm M_{m,1}(\bm H_m-\beta\bm 1^{\prime}\bm v)+\Delta \bm M_{m,2}+
\left(\frac{\bm Y_{m-1}}{|\bm Y_{m-1}|}-\bm v\right)\big(\bm H-\beta\bm 1^{\prime}\bm v\big) \nonumber\\
&+\left(\frac{\bm Y_{m-1}}{|\bm Y_{m-1}|}-\bm v\right)\big(\bm H_m-\bm H)+\bm v(\bm H_m-\bm H) \nonumber\\
\approx&\beta \bm v+ \Delta \bm M_{m,1}(\bm H_m-\beta\bm 1^{\prime}\bm v) +\Delta \bm M_{m,2}
+\left(\frac{\bm Y_{m-1}}{\beta (m-1)}-\bm v\right)\overline{\bm H}+\bm v(\bm H_m-\bm H)\nonumber
\end{align}
If we ignore the remainder $\bm v(\bm H_m-\bm H)$ and replace the martingales $\bm M_{n,1}$ and $\bm M_{n,2}$  by two (independent) Brownian motions $\bm B_1(n)\bm\Sigma_1^{1/2}$ and $\bm B_2(n)\bm\Sigma_2^{1/2}$, we conclude that
\begin{equation}\label{eqUrnapp1}\bm Y_n-\beta n\bm v\approx \bm G_1(n)\overline{\bm H}+\bm G_2(n)
\end{equation}
with $\bm G_j(t)$ satisfying the stochastic differential equation (SDE):
\begin{equation}\label{SDEGPU} d \bm G_j(t)=d \bm B_j(t) \bm\Sigma_j^{1/2}+ \frac{\bm G_j(t)}{ t}\frac{\overline{\bm H}}{\beta}dt ,\;\;  \bm G_j(0)=\bm 0 \text{ or } \bm G_j(1)=\bm 0,
\end{equation}
where $\bm B_1$ and $\bm B_2$ are two independent standard Brownian motions,  $\bm \Sigma_1=diag(\bm v)-\bm v^{\prime}\bm v$ and
 $\bm\Sigma_2=\lim\limits_{m\to \infty}\big(\sum_{k=1}^K v_k \Var\{\bm D_m^{(k)}|\mathscr{F}_{m-1}\}\big)$ and $\bm D_m^{(k)}$ is the $k$ row of $\bm D_m$.
 Further,
\begin{align*}
 & \Delta(\bm N_t-t\bm v)=  \Delta\bm M_{t,1}+\frac{\bm Y_{t-1}}{|\bm Y_{t-1}|}-\bm v\\
 \approx & d \bm B_1(t)\bm\Sigma_1^{1/2} +\frac{\bm G_1(t)\overline{\bm H}+\bm G_2(t)}{\beta t}dt
 =   d \bm G_1(t) +\frac{\bm G_2(t)}{\beta t}dt,
\end{align*}
So,
\begin{equation}\label{eqUrnapp2} \bm N_n-n\bm v\approx \bm G_1(n)+\int_{\ast}^n \frac{\bm G_2(t)}{\beta t}dt.
\end{equation}
By calculating the  variance of the Gaussian processes $\bm G_1(t)$ and $\bm G_2(t)$, one can derive the asymptotic normality of $\bm Y_n$ and $\bm N_n$.

 For details on the asymptotic normalities of urn models, one can refer to Bai, Hu and Rosneberger (2002), Hu and Zhang (2004b), Bai and Hu (2005), Zhang and Hu (2009), Zhang (2012) etc.  Among these studies, Bai, Hu and Zhang (2002), Hu and Zhang (2004b), Zhang and Hu (2009), Zhang (2012) studied  the asymptotic properties via the Gaussian approximation. The following Theorem \ref{thGPU} and Corollary \ref{CorRPW} give a summary of theses results.
\begin{theorem}\label{thGPU} Suppose $\sup_m\ep\|\bm D_m\|^{2+\delta}<\infty$. Let $\lambda_1=\beta$, $\lambda_2,\ldots, \lambda_K$  be the eigenvalues of $\bm H$, and $\lambda=\max\{Re(\lambda_2)/\lambda_1,\ldots, Re(\lambda_K)/\lambda_1\}$.
\begin{description}
  \item[(i)] If $\lambda<1$ and $\sum_{m=1}^n \|\bm H_m-\bm H\|=o(n)$ a.s., then
  $$ \frac{N_{n,k}}{n}\to \beta v_k\;\; a.s. \text{ and } \frac{Y_{n,k}}{Y_{n,1}+\cdots Y_{n,K}}\to v_k \;\; a.s. $$
  \item[(ii)] If $\lambda<1/2$ and
  \begin{equation}
  \label{eqUrnCondition} \sum_{m=1}^n \|\bm H_m-\bm H\|=o(\sqrt{n}) \;\; a.s.,
  \end{equation}
   then
  $$ \sqrt{n}\Big(\frac{\bm Y_n}{n}-\beta\bm v\Big)\overset{\mathcal D}\to N(\bm 0,\bm \Gamma_1) \text{ and } \sqrt{n}\Big(\frac{\bm N_n}{n}-\bm v\Big)\overset{\mathcal D}\to   N(\bm 0,\bm \Gamma_2). $$
  \item[(iii)] If $\lambda=1/2$ and $\sum_{m=1}^n\|\bm H_m-\bm H\|/\sqrt{m}=o(\sqrt{\log n})$ a.s., then
  $$ \frac{\sqrt{n}}{(\log n)^{\nu-1/2}}\Big(\frac{\bm Y_n}{n}-\beta\bm v\Big)\overset{\mathcal D}\to N(\bm 0,\bm \Gamma_1^{\ast}) \text{ and } \frac{\sqrt{n}}{(\log n)^{\nu-1/2}}\Big(\frac{\bm N_n}{n}-\bm v\Big)\overset{\mathcal D}\to N(\bm 0,\bm \Gamma_2^{\ast}). $$
   \item[(iv)] If $1/2<\lambda<1$ and $\sum_{m=1}^n\|\bm H_m-\bm H\|=o(n^{\lambda-\epsilon})$ a.s., then
  $$ \|\bm Y_n-\beta n\bm v\|=O(n^{\lambda}\log^{\nu-1} n) \;  a.s. \text{ and } \;\;  \|\bm N_n-n\bm v\|=O(n^{\lambda}\log^{\nu-1} n) \;\; a.s. $$
 \end{description}
 Here $\nu$ is the largest order of the Jordan blocks with respect to the eigenvalues $\lambda_s$ with $Re(\lambda_s)/\lambda_1=\lambda$.
\end{theorem}

For the randomized play-the-winner rule, we have $\beta=1$, $\lambda=p_1+p_2-1$. The problem is reduced to the one-dimensional problem:
$Y_{n,1}-nv_1\approx G_1(n)\lambda+G_2(n)$, $N_{n,1}-nv_1\approx   G_1(n)+\int_{\ast}^n \frac{G_2(t)}{t}dt$,
$$ d G_j (t)=\sigma_jd B_j(t)+\lambda \frac{G_j (t)}{t} dt $$
with $\sigma_1^2=v_1v_2=\frac{q_1q_2}{(q_1+q_2)^2}$ and $\sigma_2^2=v_1p_1q_1+v_2p_2q_2=\frac{q_1q_2(p_1+p_2)}{q_1+q_2}$.
\begin{corollary}\label{CorRPW}
\begin{description}
  \item[(a)] If $p_1+p_2<1.5$, then $\sqrt{n}\big(N_{n,1}/n-q_2/(q_1+q_2)\big)\overset{\mathcal D}\to N(0,\sigma^2_{RPW})$, where
  $$\sigma^2_{RPW}=\frac{q_1q_2[5-2(q_1+q_2)]}{[2(q_1+q_2)-1](q_1+q_2)^2},\;\; \text{ and } $$
  $$ \sqrt{n}\Big(\frac{Y_{n,1}}{Y_{n,1}+Y_{n,2}}-\frac{q_2}{q_1+q_2}\Big)\overset{\mathcal D}\to N\Big(0,\frac{q_1q_2}{(2(q_1+q_2)-1)(q_1+q_2)^2}\Big). $$
  \item[(b)]  If $p_1+p_2=1.5$ , then
  $$  \sqrt{n/\log n}\big(Y_{n,1}/n-q_2/(q_1+q_2)\big)\overset{\mathcal D}\to N(0, 4 q_1q_2),\;\;\text{ and } $$
  $$ \sqrt{n/\log n}\big(N_{n,1}/n-q_2/(q_1+q_2)\big)\overset{\mathcal D}\to N\Big(0,16 q_1q_2\Big).  $$
  \item[(c)]  If $p_1+p_2>1.5 $, then there exists a random variable $\xi$ such that
  $$n^{p_1+p_2-1} \big(Y_{n,1} - n q_2/(q_1+q_2)\big)\to \xi/2\;\; a.s., \text{ and } $$
  $$ n^{p_1+p_2-1} \big(N_{n,1} - n q_2/(q_1+q_2)\big)\to \xi \;\; a.s.  $$
  \end{description}
\end{corollary}

As a multi-treatment extension of the RPW rule, Wei (1979) proposed a GPU model to allocate patients, in which the urn is updated in the following way: at the $m$th stage, if a patient is assigned to treatment $k$ and the outcome is a success, then  a type $k$ ball is added to the urn, otherwise, if the treatment $k$ for a patient is a failure, then $1/(K-1)$ balls are added to the urn for  each of other $(K-1)$ types. This urn model is a homogenous urn model with $\bm H=\{h_{k,j}\}$, where $h_{k,k}=p_k$ and $h_{k,j}=q_k/(K-1)$ ($j\ne k$), and $q_k$ is the failure rate of treatment $k$,  $p_k=1-p_k$ is the success  probability. So,
$ v_k= \frac{1/q_k}{\sum_{j=1}^K (1/q_j)}$.

One important  class of the non-homogenous urn model is the sequential estimation-adjusted urn  model proposed by Zhang, Hu and Cheung (2006), in which the urn is updated according to the current response and the current estimate $\widehat{\bm \theta}_m$ of the parameter $\bm\theta$, and so $\bm H(\widehat{\bm \theta}_m)$ is a function of the current estimator. In this case, the fastest convergence rate of $\bm H_m-\bm H$ is $O_P(\sqrt{m})$  so that the condition
(\ref{eqUrnCondition}) is not satisfied and the term $\bm v(\bm H_m-\bm H)$ can not be ignored. In this case,
$$v(\bm H_m-\bm H)\approx \Big(\widehat{\bm \theta}_m-\bm \theta\Big)\frac{\partial [\bm v\bm H(\bm\theta)]}{\partial \bm \theta}
\approx \frac{\bm W_3(m)}{m} \frac{\partial [\bm v\bm H(\bm\theta)]}{\partial \bm \theta}, $$
where $\bm W_3(t)$ is another Brownian motion which is independent of $\bm B_1(\cdot)$, but may depend on $\bm B_2(\cdot)$.
The variance-covariance of $\Big(\bm B_1(t)\bm\Sigma_1^{1/2},\bm B_2(t)\bm\Sigma_2^{1/2},\bm W_3(t)\Big)$ is given in (5.23) of Zhang, Hu and Cheung (2006).

 Hereafter, for a vector function $\bm g(\bm x)$,
its derivative  $\frac{\partial g(\bm x)}{\partial\bm x}=\Big(\frac{\partial g_k(\bm x)}{\partial x_j};j,k=1,\ldots,K\Big)$ is a matrix.
In the approximation (\ref{eqUrnapp1}) and  (\ref{eqUrnapp2}), $\bm G_2(t)$ should be replaced with $\bm G_2(t)+\bm G_3(t)$ and
$$ d \bm G_3(t)= \frac{ \bm W_3(t)}{t} dt +\frac{\bm G_3(t)}{t} \frac{\overline{\bm H}}{\beta}dt. $$
The conclusions of Theorem~\ref{thGPU} remains true with a different group of variance-covariance  matrices.

In particular, suppose  $\widehat{\bm \theta}_m=(\widehat{\bm \theta}_{m,1},\ldots,\widehat{\bm \theta}_{m,K})$ is the maximum likelihood estimator of the parameter $\bm \theta=(\bm\theta_1,\ldots,\bm \theta_K)$ based on the responses available up to stage $m$. If we define
$$\bm D_m=\bm H_m=\beta\bm 1^{\prime}\bm \rho\big(\widehat{\bm \theta}_m\big) \text{ with } \rho_k(\cdot)>0 \text{ and } \sum_k \rho_k(\cdot)=1, \;\; \beta>0, $$
 then $\bm N_n-n\bm v=O(\sqrt{n\log\log n})$ a.s. and
\begin{equation}\label{eqSEU1}
\sqrt{n}\Big(\frac{\bm N_n}{n}-\bm\rho \Big)\overset{\mathcal D}\to N\Big(\bm 0, diag(\bm \rho)-\bm\rho^{\prime}\bm\rho+6\bm\Sigma_{LB}\Big),
\end{equation}
where $\bm \rho=\bm\rho(\bm\theta)$,
\begin{equation}\label{eqLB} \bm\Sigma_{LB}=\Big(\frac{\partial \bm\rho(\bm\theta)}{\partial \bm\theta}\Big)^{\prime} daig\Big(\frac{\bm I_1^{-1}(\bm\theta_1)}{\rho_1(\bm\theta)},\cdots,\frac{\bm I_K^{-1}(\bm\theta_K)}{\rho_K(\bm\theta)}\Big) \frac{\partial \bm\rho(\bm\theta)}{\partial \bm\theta},
\end{equation}
and $\bm I_k(\bm\theta_k)$ is the Fisher information for a single observation on treatment $k=1,\ldots, K$. Hereafter, $\bm I_k^{-1}$ denotes the inverse matrix
of $\bm I_k$.

When the estimator $\widehat{\bm\theta}_m$ of the parameter $\bm\theta$ is utilized to update the allocation probabilities,  the adaptive randomization scheme can start only  after an initial estimator is defined  because at the first few steps there are insufficient data for  estimating $\bm\theta$. In general, there are three ways to overcome this problem: 1) initial $Km_0$ patients ($m_0$ is some small positive integer) are randomized to treatments $1,\ldots,K$ by means of  some restricted randomization design, and their outcomes are used to estimate   $\bm\theta$;  2)   choose an initial value $\bm\theta_0$ as the estimator  until sufficient amount of data are collected to estimate $\bm\theta$ (the value $\bm\theta_0$ is usually a guess  value of the parameter or an estimate from other trials); 3)  apply the Bayesian estimation method.

\subsection{Drop-the-Loser}

The asymptotic normality for the randomized play-the-winner rule as well as its extensions to generalized P\'olya urn models
 can be obtained only when the condition $\rho\le 1/2$ is satisfied, and the variabilities are very large. Many of these designs are slow to converge and produce less powerful treatment comparison hypothesis tests. In the family of urn models, a major advance was made by
Ivanova (2003), who introduced "the {\em drop-the-loser}  rule", an urn model design with the same limiting allocation as the RPW rule but   with much lower variability.  In the drop-the-loser rule,  an urn is considered with balls of $(K+1)$ types, type $0,1,\ldots, K$, when comparing $K$ treatments. Types $1,\ldots, K$ are called treatment types, and type $0$ is called the immigration type. When a patient is ready for randomization, a ball is drawn at random. If it is of a treatment type, the corresponding treatment is assigned and the patient's response is observed.   If the response is a success, the ball is replaced and the urn remained unchanged. If a failure the ball is not replaced. When an immigration type ball is drawn, no treatment assignment is made, and the ball is return to the urn together with one ball of each treatment type. Ivanova (2003,2006) showed the asymptotic normality by embedding the urn process to a death-and-immigration process.
\begin{theorem}\label{thDL} Let
$ \bm v=(v_1,\cdots, v_K)$ with $ v_k=\frac{1/q_k}{\sum_{j=1}^K (1/q_j)}. $
 Then
$$ \sqrt{n}\Big(\frac{\bm N_n}{n}-\bm v\Big)\overset{D}\to N\big(\bm 0,\bm\Sigma_{DL}\big) $$
with
$ \bm\Sigma_{DL} = (\bm I-\bm v^{\prime}\bm 1)diag\Big( \frac{v_1p_1}{q_1},\cdots, \frac{v_Kp_K}{q_K}\Big)(\bm I- \bm 1^{\prime}\bm v). $
It can be verified that
$$ \bm\Sigma_{DL}=\Big(\frac{\partial \bm v}{\partial \bm q}\Big)^{\prime} diag\Big( \frac{p_1q_1}{v_1},\cdots, \frac{p_Kq_K}{v_K}\Big)\frac{\partial \bm v}{\partial\bm q}. $$
\end{theorem}
In particular, for the two-treatment case,
$$ \sqrt{n}\Big(\frac{N_{n,1}}{n}-\frac{q_2}{q_1+q_2}\Big) \overset{D}\to N\Big(0,\frac{q_1q_2(p_1+p_2)}{(q_1+q_2)^3}\Big). $$
The asymptotic normality of DL rule holds for all cases of $0<p_1,p_2<1$ and the variability $\frac{q_1q_2(p_1+p_2)}{(q_1+q_2)^3}$ is much smaller than the one of the RPW rule. For the multi-treatment case, the formula of $\bm \Sigma_{LD}$ in Ivanova (2006) is given in a different expression. The present one is due to Zhang et al. (2011).

\subsection{Generalized Drop-the-Loser and Immigrated Urn Model}
 Generalizations of
the drop-the-loser rule can be found in Zhang et al. (2007), Sun, Cheung and Zhang (2007) and
Zhang et al. (2011).  Zhang et al. (2011) proposed an {\em immigrated urn}  model which provides a unified theory of urn models for clinical trials. In an immigrated urn model, as in the drop-the-loser rule, the urn contains balls of $(K+1)$ types, where types $1,\ldots, K$ stand for treatment types, and type $0$ stands for the immigration type. After $m$ assignments, suppose the urn composition is $(Y_{m,0},Y_{m,1},\ldots,Y_{m,K})$. For the $(m+1)$-th patient's treatment assignment, a ball is drawn at random. If an immigration type ball is drawn, no assignment is made, and the ball is return to the urn together with $a_{m+1,k}\ge 0$ balls of type $k$, $k=1,\ldots,K$. The process is repeated until a ball of a treatment type is drawn. If the ball drawn  is of   type $k$ ($k=1,\ldots,K$), the corresponding treatment is assigned and the patient's response $\bm \xi(m+1)$ is observed.  The ball is also return with additional $D_{k,j}(m+1)$ balls of each treatment type $j$, $j=1,\ldots,K$.  $D_{k,j}(m+1)$ is a function of the response $\bm\xi(m+1)=\big(\xi_1(m+1),\ldots, \xi_K(m+1)\big)$. In the IMU model, the diagonal elements of $\bm D_{m+1}$ allow negative values, which means the  dropping of balls, and $a_{m+1,k}=a_k(\widehat{\bm\theta}_m)$ can be a function of the current estimator $\widehat{\bm\theta}_m$  of the parameter $\bm\theta$. The vector $\bm a_m=(a_{m,1},\ldots, a_{m,K})$ are called the immigration rates.  When $a_{m,k}\equiv 0$ and $Y_{0,0}=0$, the IMU   reduces to GPU. When $\bm D_m$ is diagonal and its elements have negative means, the IMU is the generalized drop-the-loser  rule proposed by Zhang et al. (2007) and Sun, Cheung and Zhang (2007). Here when the urn allows balls with negative numbers, we assume that the  balls of a type with a negative number have no-chance to be selected and so the selection probabilities are the proportions of positive numbers of balls in the urn.

For considering the theory of the IMU model, without loss of generality,  we suppose  the parameter $ \theta_k$ is one-dimensional  and  the mean of the response $\xi_k(m)$ and the estimator $\widehat{\theta}_{m,k}$ is  the current sample mean, and assume that $\{(\bm D_m,\bm\xi(m))\}$ are   i.i.d. Let $\bm H=\ep[\bm D_m]$ be
the mean matrix as in the GPU model. If $\bm H\bm 1^{\prime}=\beta \bm 1^{\prime}$ with $\beta>0$, i.e., at each stage the average of the total number of  balls added according the treatments are positive, then
the total number of balls in the urn gradually increases to infinity.
Hence, the probability of drawing an immigration ball drops to zero. For this
case,   the IMU model is asymptotically equivalent to the
generalized P\'olya urn model without immigration,  and the conclusions of Theorem~\ref{thGPU} remain true.
When $\bm H \bm 1^{\prime}<\bm 0^{\prime}$, the urn composition is mainly updated by the immigration, and we have the following theorem.
\begin{theorem} Let $\bm A=(-\bm H)^{-1}(\bm I-\bm 1^{\prime}\bm v)$,
\begin{eqnarray*}
&\bm v(\bm\theta)=(v_1,\ldots,v_K)=\frac{\bm a(\bm \theta)(-\bm H)^{-1}}{\bm a(\bm \theta)(-\bm H)^{-1}\bm 1^{\prime}},&\\
&\bm \Sigma_{11}=\sum_{k=1}^K v_k \Var\{\bm D_1^{(k)}\},\;\;\; \bm\Sigma_D=\bm A^{\prime}\bm\Sigma_{11}
\bm A,&\\
& \bm
\Sigma_{12}=(\Cov\{D_{1,kj},\xi_k\}; j,k=1,\ldots, K), \;\;\; \bm\Sigma_{D\xi}=\bm A^{\prime} \bm\Sigma_{12}  \frac{\partial \bm v(\bm\theta)}{\partial\bm\theta},&\\
&\bm\Sigma_{\xi}=\Big(\frac{\partial \bm v(\bm\theta
)}{\partial\bm\theta}\Big)^{\prime} diag\Big(\frac{\Var\{\xi_{1,1}\}}{v_1},\ldots,\frac{\Var\{\xi_{1,K}\}}{ v_K}\Big)\frac{\partial
\bm v(\bm\theta)}{\partial\bm\theta},&
\end{eqnarray*}
and $\bm\Sigma=\bm\Sigma_{D}+2\bm\Sigma_{\xi}+\bm\Sigma_{D\xi}+\bm\Sigma_{D\xi}^{\prime}$. Suppose
$\ep\|\bm D_m\|^{2+\epsilon}<\infty$,  $\ep\|\bm \xi(m)\|^{2+\epsilon}<\infty$, and  $\bm H\bm 1^{\prime}<\bm 0^{\prime}$,   $Y_{0,0}>0$.  Then
 $$Y_{n,k}=o(n^{1/2-\epsilon})\;\; a.s., \;\; k=1,\ldots, K, \;\; \bm N_n-n\bm v=O(\sqrt{n\log\log n})\; a.s.$$
 $$  \; \;  \text{ and } \;
\sqrt{n}\Big(\frac{\bm N_n}{n}-\bm v(\bm\theta)\Big)\overset{ D }\to N(\bm
0,\bm\Sigma).
$$
In particular,
\begin{description}
  \item[(i)]  when $\bm D_m\equiv const$ (for example $\bm D_m=diag(-1,\ldots,-1)$), one has
\begin{equation}\label{eqGDL3}\sqrt{n}\Big(\frac{\bm N_n}{n}-\bm v(\bm\theta)\Big)\overset{ D }\to N(\bm
0,2\bm\Sigma_{\xi});
\end{equation}
  \item[(ii)] when $\bm a(\bm\theta)\equiv const$ and each $D_{m,kj}$ is a linear function  of $\xi_{m,k}$, $j=1,\ldots,K$, so that $\bm v$ is a function of
$\theta_k=\ep\xi_{m,k}$,$k=1,\ldots,k$, one has
$$\sqrt{n}\Big(\frac{\bm N_n}{n}-\bm v(\bm\theta)\Big)\overset{ D }\to N(\bm
0,\bm\Sigma_{\xi}).$$
\end{description}
 \end{theorem}

 When $\bm H\bm 1^{\prime}=\bm 0$, we have $\bm N_n-n\bm v=O(\sqrt{n\log\log n})$ a.s. and $\bm N_n-n\bm v=O(\sqrt{n})$ in probability. The asymptotic normality is still an open problem.

\subsection{Randomly Reinforced Urn}
Another type of urn models, called {\em randomly reinforced urn} (RRU), are also proposed for randomizing patients to treatments. In a response-adaptive design driven by a RRU model, an observation of an outcome from treatment $k$  results in only adding balls of  the same type $k$. A RRU procedure leads to an extreme limiting allocation so that the sample allocation proportions of the best treatment converges to $1$, and others converge to zero.    For the properties of the RRU design, one may refer to Li, Durham  and Flournoy (1996), May and Flournoy (2009), Zhang et al. (2014a) etc.
However, the RRU design has very high  variability and so is not strongly recommended in the term of power.

\section{Optimal  and Efficient RAR Designs} \label{sectOpEffic}
\setcounter{equation}{0}

The RPW rule and many of its extensions based on urn models were  proposed by the intuitive motivation of placing more patients on the superior treatment, and they were not designed to optimize any criterion.  Hu and Rosenberger (2003) formalized the development of optimal response-adaptive randomization procedures using the following three steps:
\begin{description}
  \item[\rm 1)] An
optimal allocation is derived as a solution to some formal optimization problem.
 \item[\rm 2)] A response-adaptive randomization procedure is chosen to converge to the
optimal target. The procedure should be fully randomized, have minimal variability,
and high speed of convergence to the chosen optimal allocation.
\item[\rm 3)] Operating characteristics of the chosen response-adaptive randomization
design are studied theoretically and by simulation under a variety of standard
to worst-case scenarios.
 \end{description}

\subsection{Optimization}\label{subsectOpt}
For two-treatment trials, a general optimization problem is described in Jennison
and Turnbull (2000), which led to the development of optimal response-adaptive
randomization designs for trials with binary responses (Rosenberger et al. (2001);
Ivanova and Rosenberger (2001); Rosenberger and Hu (2004); Baldi Antognini
and Giovagnoli (2010)), normal outcomes (Biswas and Mandal (2004); Zhang and
Rosenberger (2006); Gwise, Hu and Hu (2008); Biswas and Bhattacharya (2009, 2010,
2011)), and survival outcomes (Zhang and Rosenberger (2007)). The most interesting optimal allocation targets are the Neyman allocation for continuous response and the RSIHR allocation for binary responses.

In Jennison and Turnbull (2000)'s approach, let $\xi_{m,1}$ arise from a   $N(\mu_1,\sigma_1^2)$ distribution and   $\xi_{m,2}$ arise from a   $N(\mu_2,\sigma_2^2)$ distribution, $m=1,2,\ldots$. For testing the treatment effect $\theta=:\mu_1-\mu_2=0$, a natural test is the Wald test given by
$$ Z=\frac{\widehat{\mu}_1-\widehat{\mu}_2}{\sqrt{\frac{\widehat{\sigma}_1^2}{n_1}+ \frac{\widehat{\sigma}_2^2}{n_2}}},$$
where $n_1$ and $n_2$ are the sample size for treatment $1$ and $2$. The (asymptotic) power of the test is a decreasing of
$$\eta(\theta)= \frac{ \sigma_1^2}{n_1}+ \frac{ \sigma_2^2}{n_2}. $$
Fixing $\eta$ to a constant, say $\delta$, we wish to find the value of $\frac{n_1}{n_1+n_2}$ that minimizes
$$ u(\theta)n_1+v(\theta)n_2, $$
where $u(\cdot)$ and $v(\cdot)$ are appropriately chosen functions of $\theta$. Because we wish to put more patients on treatment $1$ if $\theta>0$ and more patients on treatment $2$ if $\theta<0$, Jennison and Turnbull explore functions where $u(\cdot)$ and $v(\cdot)$ are strictly positive, and $u(\theta)$ is decreasing in $\theta$ for $\theta<0$ and $v(\theta)$ is increasing in $\theta$ for $\theta>0$.  By using the Lagrange multiplier method and minimizing
$$  u(\theta)n_1+v(\theta)n_2+\lambda\Big(\frac{ \sigma_1^2}{n_1}+ \frac{ \sigma_2^2}{n_2}-K\Big), $$
  a  minimum is achieved at
$$ \frac{n_1}{n_1+n_2}= \rho_1=:\frac{\sigma_1/\sqrt{u(\theta)}}{\sigma_1/\sqrt{u(\theta)}+\sigma_2/\sqrt{v(\theta)}}. $$
If $u(\cdot)\equiv v(\cdot)\equiv 1$, then we have $\rho_1=\sigma_1/(\sigma_1+\sigma_2)$, which is   the Neyman allocation. This allocation maximizes the power of the usual $Z$-test for fixed sample size $n_1+n_2=n$.

In the case of  the binary responses, for testing the  equality of treatment effects,  the Wald test is given by
$$ Z=\frac{\widehat{p}_1-\widehat{p}_2}{\sqrt{\frac{\widehat{p}_1\widehat{q}_1}{n_1}+ \frac{\widehat{p}_2\widehat{q}_2}{n_2}}},$$
where $\widehat{p}_k$ is the estimator of the cured rate $p_k$, and $\widehat{q}_k =1-\widehat{p}_k$, $k=1,2$. Rosenberger et al. (2001) suggested  fixing the asymptotic variance  $\eta(q_1,q_2)= \frac{ p_1q_1}{n_1}+ \frac{p_2q_2}{n_2}$ to a constant, say $\delta$, and minimizing the average failure number $q_1n_1+q_2n_2$. The  minimum is achieved   at
$$ \frac{n_1}{n_1+n_2}= \rho_1=:\frac{\sqrt{p_1} }{\sqrt{p_1}+\sqrt{p_2}}. $$
We refer to this as RSIHR allocation.

Many other optimal allocations    can be found in literature. For example, Zhang and Rosenberger (2006) proposed the following allocation for normal responses with positive means by minimizing the mean total response for fixed power:
$$ \rho_1(\bm\theta)=\frac{ \sqrt{\mu_2}\sigma_1}{ \sqrt{\mu_2}\sigma_1+ \sqrt{\mu_1}\sigma_2}; $$
By minimizing  the total number of patients with normal response greater than a constant given  $c$, Biswas and Mandal (2004)  obtained an allocation as
$$  \rho_1(\bm\theta)
=\frac{\sqrt{\Phi\big(\frac{u_2-c}{\sigma_2}\big)}\sigma_1}{\sqrt{\Phi\big(\frac{u_2-c}{\sigma_2}\big)}\sigma_1
+\sqrt{\Phi\big(\frac{u_1-c}{\sigma_1}\big)}\sigma_1}. $$

For multi-arm clinical trials,
Tymofyeyev,  Rosenberger and Hu (2007) introduced a general approach for finding allocations to minimize a weighted
sum of treatment sample sizes subject to minimal constraints on the power of a
homogeneity test and treatment proportions in the trial. Let $\phi(n_1,\ldots,n_K)$ be the noncentrality parameter of a suitable multivariate test statistics of interest under the alterative hypothesis, where $n_k$ is sample size for treatment $k$, $n_1+\cdots n_K=n$. We assume that the noncerntrality parameter is a concave function with nonnegative gradient.
Tymofyeyev, Rosenberger and Hu (2007) suggested  considering a general  optimization problem:
\begin{align*}
  \min_{n_1,\ldots,n_K} \;\; & \sum_{j=1}^K w_j n_j, \\
   \text{ such that }\;\;   & \frac{n_k}{\sum_{j=1}^K n_j}\ge B, \;\; k=1,\ldots,K,\\
   & \phi(n_1,\ldots,n_K)\ge C,
\end{align*}
or equivenlently,
\begin{align*}
  \max_{m_1,\ldots,m_K} \;\; &  \phi(m_1,\ldots,m_K)  \\
   \text{ such that }\;\;   & \frac{m_k}{\sum_{j=1}^K m_j}\ge B, \;\; k=1,\ldots,K,\\
   & \sum_{j=1}^K w_j m_j\le M,
\end{align*}
to derive the allocation target $\rho_k=n_k/\sum_{j=1}^K n_j=m_k/\sum_{j=1}^K m_j$.
Here $w_j$ are positive weights, and the nonnegative constant $B$ with $KB\le 1$ is a lower bound for the proportion $n_k/\sum_{j=1}^K n_j$ (resp. $m_k/\sum_{j=1}^K m_j$) that allows us to control explicitly the feasible region of the problem.
Under this framework,  Jeon and Hu (2010), for binary trials with $K = 3$ treatments, obtained the analytic solution for an allocation
minimizing expected treatment failures in the trial subject to minimal constraints
on power and treatment proportions. Similar results were obtained by Zhu and
Hu (2009) for exponential outcomes and by Sverdlov, Tymofyeyev and Wong  (2011) for censored
exponential outcomes.

Also, there  is a lot papers considering the $D$-optimization, $DA$-optimization etc. In general, let $\bm M(\bm\rho,\bm\theta)$ denote the Fisher information matrix for $\bm\theta$ given a design allocation $\bm\rho=(\rho_1,\ldots,\rho_K)$. By minimizing $\bm M^{-1}(\bm\rho,\bm\theta)$ in some sense (by choice of $\bm\rho$) one can achieve most accurate inference for the parameters of interest.   For normal response trials with heteroscedastic outcomes,
Wong and Zhu (2008) and Gwise, Zhu and Hu (2011) obtained $DA$-optimal designs that
maximize efficiency for estimating treatment contrasts. Sverdlov and Rosenberger (2013a) give a comprehensive overview of various single- and multiple-objective optimal allocation designs that are available in the literature.

\subsection{Target Driven Randomization}\label{subsectRand}
\setcounter{equation}{0}

In general, the optimal allocation  proportion vector $\bm \rho =(\rho_1,\ldots, \rho_K)$ derived as a solution to some   optimization problem is a function of the distribution parameters $\bm\theta=(\bm\theta_1,\ldots, \bm\theta_K)$. In practice, one can construct a response-adaptive randomization design such that the sample allocation proportions sequentially converge  to the chosen optimal allocation $\bm\rho(\bm\theta)=\big(\rho_1(\bm\theta),\ldots,\rho_K(\bm\theta)\big)$. We shall assume that the target function $\rho(\bm\theta)$ is a continuous function on the parameter space and twice differentiable at the true value of the parameter $\bm\theta=(\bm\theta_1,\ldots, \bm\theta_K)$. In general, one can use a smoothing mothod to modify the target function (c.f. Tymofyeyev, Rosenberger and Hu  (2007)).

In the two-treatment trails with binary responses, the sample allocations $N_{n,1}/n$ and $N_{n,2}/n$  in both the RPW rule and DL rule  sequentially converge  to  allocation proportion $\rho_1=q_2/(q_1+q_2)$ and $\rho_2=1-\rho_1$.   However, the RPW and the DL rule can only target this specified allocation.

In general, given an allocation proportion vecotr $\bm\rho(\bm\theta)$, if we apply the SEU model   with adding rules  $\bm D_m=\beta \bm 1^{\prime}\bm \rho\big(\widehat{\bm \theta}_m\big)$ ($\beta>0$), then we will have $\bm N_n/n\to \bm \rho(\bm\theta)$ a.s. and the asymptotic normality given by  (\ref{eqSEU1}). Here we always assume that  $\widehat{\bm\theta}_m$ is the MLE of $\bm\theta$, or simply  the sample means of the responses, based on the data from the previous $m$ trials. In a SEU model   with adding rules  $\bm D_m=\beta \bm 1^{\prime}\bm \rho\big(\widehat{\bm \theta}_m\big)$, when the $(m+1)$-th patient is randomized,   no matter what its response is, the urn is updated by adding additional  $\beta \rho_j(\widehat{\bm\theta}_m)$ balls of type $j$, $j=1,\ldots ,K$. The outcomes of treatments are only used to get the estimator $\widehat{\bm\theta}_m$.

 Also, if we apply the IMU model with immigration rates $\bm a(\widehat{\bm\theta}_m)=\beta\bm \rho(\widehat{\bm\theta}_m)$ ($\beta>0$) and adding rules $\bm D_m=diag(-1,\cdots,-1)$, then we will also have $\bm N_n/n\to \bm \rho(\bm\theta)$ a.s. and the asymptotic normality given by  (\ref{eqGDL3}).
 This procedure  is the   generalized drop-the-loser rule  proposed by Zhang et al. (2007) and  Sun, Cheung and Zhang (2007).  In the GDL rule, when a treatment ball is drawn, it is always dropped. When an immigration ball is drawn, the ball is returned  with additional $\beta\rho_j(\widehat{\bm \theta}_m)$ balls of type $j$, $j=1,\ldots, K$. The outcomes of treatments are only used to estimate the parameter $\bm\theta$ and get   the current estimated immigration rates $\bm\rho(\widehat{\bm\theta}_m)$.

Another simple approach to  construct a response-adaptive randomization design such that $\bm N_n/n\to \bm \rho(\bm\theta)$ is defining the allocation probabilities of the $(m+1)$-th patient by
$$ p_{m+1,k}= \widehat{\rho}_{m,k}, \;\; k=1,\ldots, K, $$
where $\widehat{\bm \rho}_m=\big(\widehat{\rho}_{m,1},\ldots, \widehat{\rho}_{m,K}\big)$ with $\widehat{\rho}_{m,k}=\rho_k\big(\widehat{\bm\theta}_m\big)$ denotes the estimated target allocation.
 This is the {\em sequential maximum likelihood procedure} (SMLP) proposed by Melfi and Page (2000) and Melfi, Page, and Geraldes (2001).

 A general allocation rule is defining   the allocation probabilities as a function of both current sample allocation proportions and  the estimated target allocation:
$$  p_{m+1,k}= g_k\left(\bm N_m/m, \widehat{\bm \rho}_m\right), \;\; k=1,\ldots, K,$$
where $\bm g(\bm x,\bm y)=\big(g_1(\bm x,\bm y),\ldots, g_K(\bm x,\bm y)\big)$ ($g_k(\bm x,\bm y)\ge 0$ and $\sum_{k=1}^Kg_k(\bm x,\bm y)=1$)  is called the allocation function. This is the {\em doubly adaptive biased coin design}  proposed by Hu and Zhang (2004a), extending the work of Eisele (1994).

For a general function $\bm g$, the convergence of $\bm  N_n/n$ is related to the stability of the following ordinary differential equation:
$$ \dot{\bm x} =\bm x-\bm g\big(\bm x, \bm\rho(\bm\theta)\big), \text{ with } \bm x=\bm x(s), $$
where $\dot{\bm x}(s)$ is the derivative of $\bm x(s)$ (c.f. Zhang (2014)).  If the function $\bm g$ is chosen such that $g_k(\bm x,\bm y)\le \lambda (x_k-y_k)$ whenever $x_k>y_k$, where $0\le \lambda<1$, then
$$ \frac{\bm N_n}{n}\to \bm\rho(\bm\theta) \;\; a.s. $$
This is proved by Hu and Zhang (2004a) and Hu et al. (2008), and they proposed the following allocation function:
\begin{equation} \label{eqDBCDfunction} g_k(\bm x,\bm y)=\frac{y_k\big(\frac{y_k}{x_k}\big)^{\gamma}}{\sum_{j=1}^K y_j\big(\frac{y_j}{x_j}\big)^{\gamma}}, \;\; k=1,\ldots, K,
\end{equation}
where $\gamma\ge 0$ is user-defined parameter controlling the degree of randomness ($\gamma=0$ is almost completely randomized and $\gamma=\infty$ is almost deterministic procedure). The SMLP is a special case of DBCD with $\gamma=0$.

  For considering the asymptotic normality, we note that
$$ \widehat{\bm\theta}_{m,k}-\bm\theta_k\approx \frac{\sum_{j=1}^m X_{j,k}\bm\eta_{m,k}}{N_{m,k}} \approx \frac{\bm B_k(m) \bm I_k^{-1/2}\sqrt{\rho_k}}{m\rho_k}=\frac{\bm B_k(m) \bm I_k^{-1/2}}{m\sqrt{\rho_k}},
$$
$$ \bm M_{n}=\sum_{j=1}^m (\bm X_j-\ep[\bm X_j|\mathscr{F}_{j-1}])\approx \bm B(t)\bm\Sigma_1^{1/2}, $$
where $\bm \eta_{m,k}$ is a function of the response $\bm \xi(m)$ with  $\ep\bm\eta_{m,k}=\bm 0$ and $\Var(\bm\eta_{m,k})=\bm I_k^{-1}$ with
$\bm I_k=\bm I_k(\bm\theta_k)$ being  the Fisher information for a single observation on treatment $k=1,\ldots, K$, $\bm B(t), \bm B_1(t),\ldots, \bm B_K(t)$  are independent  multi-dimensional standard Browian motions, $\bm\Sigma_1=diag(\bm\rho)-\bm\rho^{\prime}\bm\rho$.
Hereafter, for a symmetric and positive definite matrix $\bm \Sigma$,  $\bm\Sigma^{1/2}$ is the symmetric matrix such that $\bm\Sigma^{1/2}\bm\Sigma^{1/2}=\bm\Sigma$, and $\bm\Sigma^{-1/2}$ is the inverse matrix of $\bm\Sigma^{1/2}$ satisfying $\bm\Sigma^{-1/2}\bm\Sigma^{-1/2}=\bm\Sigma^{-1}$.
Write $\bm W(t)=\big(\bm B_1(t),\ldots, \bm B_K(t)\big)$ and $\bm I(\bm\theta)=daig\big(\rho_1(\bm\theta)\bm I_1,\ldots, \rho_k(\bm \theta)\bm I_K\big)$. Then
\begin{align*}
\widehat{\bm\theta}_m-\bm\theta\approx & \frac{\bm W(m)\bm I^{-1/2}(\bm\theta)}{m}, \\
\Delta(\bm N_m-m\bm v)\approx&\Delta \bm M_{m,1}+\Big(\frac{\bm N_{m-1}}{m-1}-\bm v\Big)\frac{\partial \bm g(\bm \rho,\bm\rho)}{\partial \bm x}
+\Big(\widehat{\bm\theta}_{m-1}-\bm\theta\Big)\frac{\partial \bm\rho(\bm\theta)}{\partial \bm\theta}\frac{\partial \bm g(\bm \rho,\bm\rho)}{\partial \bm x}.
\end{align*}
So
$$\bm N_n-n\bm v\approx \bm G(n) \;\; \text{ with } \bm G \text{ satisfying the SDE:}$$
$$ d \bm G(t)= d\bm B(t)\bm\Sigma_1^{1/2} -\frac{\bm G(t)}{t}   \frac{\partial \bm g(\bm \rho,\bm\rho)}{\partial \bm x} dt +\frac{\bm W(t)}{t}  \bm I^{-1/2}(\bm\theta)\frac{\partial \bm\rho(\bm\theta)}{\partial \bm\theta}\frac{\partial \bm g(\bm \rho,\bm\rho)}{\partial \bm x}dt. $$
Suppose the allocation function is chosen as in (\ref{eqDBCDfunction}), then $  \frac{\partial \bm g(\bm \rho,\bm\rho)}{\partial \bm x}=-\gamma (\bm I-\bm 1^{\prime}\bm \rho)$ and $  \frac{\partial \bm g(\bm \rho,\bm\rho)}{\partial \bm y}=(\gamma+1) (\bm I-\bm 1^{\prime}\bm \rho).$ The SDE can be simplified to
  $$ d \bm G(t)= d\bm B(t)\bm\Sigma_1^{1/2} -\gamma \frac{\bm G(t)}{t}  dt  +(\gamma+1)\frac{\bm W(t)}{t}  \bm I^{-1/2}(\bm\theta)\frac{\partial \bm\rho(\bm\theta)}{\partial \bm\theta} dt, $$
which has a solution
  $$ \bm G(t)=t^{-\gamma}\int_0^t x^{\gamma} d \bm B(x)\bm\Sigma_1^{1/2}+(\gamma+1)t^{-\gamma} \int_0^tx^{\gamma-1}\bm W(x)dx  \bm I^{-1/2}(\bm\theta)\frac{\partial \bm\rho(\bm\theta)}{\partial \bm\theta}. $$
  By deriving the variability of the Gaussian process, we conclude the asymptotic normality.
\begin{theorem}\label{thDBCD} Suppose the distributions $f_1(\cdot|\bm\theta_1),\ldots,f_K(\cdot|\bm\theta_K)$ of the outcomes of treatment $k=1,\ldots, K$ follow an exponential family. Let $\bm g(\bm x,\bm y)$ be defined as (\ref{eqDBCDfunction}). Then
$$ \sqrt{n}(\widehat{\bm\theta}_n-\bm\theta)\overset{D}\to N\Big(\bm 0,\bm I^{-1}(\bm\theta)\Big), $$
$$ \bm N_n-n\bm \rho=O(\sqrt{n\log\log n})\; a.s. \text{ and } \sqrt{n}\Big(\frac{\bm N_n}{n}-\bm\rho\Big)\overset{D}\to N\Big(\bm 0,\bm\Sigma_{\gamma}\Big), $$
where
$$\bm\Sigma_{\gamma}=\bm\Sigma_{LB}+\frac{1}{1+2\gamma}\Big(diag(\bm \rho)-\bm\rho^{\prime}\bm\rho+\bm\Sigma_{LB}\Big) $$
with $\bm\Sigma_{LB}$ being defined as in (\ref{eqLB}).
\end{theorem}
We can also derive the asymptotic joint normality of $\widehat{\bm\theta}_n$ and $\frac{\bm N_n}{n}$, the asymptotic co-variance of them is
$\Cov\big\{\widehat{\bm\theta}_n,\frac{\bm N_n}{n}\big\}\sim \frac{1}{n}\bm I(\bm\theta)\frac{\partial \bm\rho(\bm\theta)}{\partial \bm\theta}$.

The asymptotic variability  $\bm\Sigma_{\gamma}$ achieves its largest value when $\gamma=0$, and approaches its minimum value $\bm\Sigma_{LB}$   as $\gamma\to \infty$.

\bigskip
In the case of two-treatment trials with binary responses, the RPW rule, the DL rule, the SEU procedure   with $\bm D_m=\bm 1^{\prime}\bm \rho(\widehat{p}_{m,1},\widehat{p}_{m,2})$, the GDL rule, the SMLP and the DBCD can all applied to target the  allocation $q_2/(q_1+q_2)$. The asymptotic variabilities of the sample allocation proportions $N_{n,1}/n$ (after normalized by $\sqrt{n}$)  are given in Table~\ref{tab1}.
\begin{table}[h]\begin{center} \renewcommand{\arraystretch}{1.3}
\vspace{0.2cm}
\begin{tabular}{|c|c|c|c|}
  \hline
  Design & RPW & DL &  SMLP   \\
   Variability ($\sigma^2$) & $\begin{matrix} \frac{q_1q_2[3+2(p_1+p_2)]}{[2(q_1+q_2)-1](q_1+q_2)^2}, & q_q+q_2>\frac{1}{2}, \\
   \infty, & q_1+q_2\le \frac{1}{2}\end{matrix}$  & $\frac{q_1q_2(p_1+p_2)}{(q_1+q_2)^3}$ &   $\frac{q_1q_2(2+ p_1+p_2) }{ (q_1+q_2)^3}$  \\
  \hline
   Design  & SEU & GDL & DBCD \\
   Variability ($\sigma^2$)   & $\frac{q_1q_2[2+5(p_1+p_2)]}{(q_1+q_2)^3}$ & $\frac{2q_1q_2(p_1+p_2)}{(q_1+q_2)^3}$   & $\frac{q_1q_2\left[2+(1+2\gamma)(p_1+p_2)\right]}{(1+2\gamma)(q_1+q_2)^3}$\\
  \hline
\end{tabular}\end{center}
\caption{The asymptotic variabilities of RAR procedures with the same traget $\frac{q_2}{q_1+q_2}$. }
\label{tab1}\end{table}
It can be verified  that  the RPW rule always has the largest variability and the DL rule has the smallest variability. In fact,
$$ \sigma^2_{RPW}>\sigma^2_{SEU}>\sigma^2_{SMLP}> \begin{matrix} \sigma^2_{GDL} \\ \sigma^2_{DBCD}\end{matrix} >\sigma^2_{DL}, $$
for all $0<q_1,q_2<1$.

The RPW rule and DL rule can only target the allocation $q_2/(q_1+q_2)$. The SEU design with $\bm D_m=\bm 1^{\prime}\bm \rho(\widehat{\bm\theta}_m)$, the GDL, the SMLP and the DBCD can be used to target any desired allocation.   Table~\ref{tab2} gives their asymptotic variabilities.
\begin{table}[h]\begin{center} \renewcommand{\arraystretch}{1.3}
\vspace{0.2cm}
\begin{tabular}{|c|c|c|c|c|}
  \hline
  Design  & SEU& GDL & SMLP& DBCD \\
   Variability ($\bm \Sigma$)  & $\bm\Sigma_1+6\bm\Sigma_{LB}$ & $2\bm\Sigma_{LB}$ & $\bm\Sigma_1+2\bm\Sigma_{LB}$ & $\frac{1}{1+2\gamma}\bm\Sigma_1+  \frac{2+2\gamma}{1+2\gamma}\bm\Sigma_{LB}$\\
  \hline
\end{tabular}\end{center}
\caption{The asymptotic variabilities of RAR procedures with the same traget $\bm\rho(\bm\theta)$. }
\label{tab2}\end{table}

\subsection{Variability and Efficiency}
For assessing theoretical  operating characteristics of candidate designs, the first step is assessing the theoretical (limit) allocation proportions under   certain criteria, which  is the optimization problem as we discuss in subsection~\ref{subsectOpt}.  The allocation proportion of the RPW rule is not optimal under usual criteria. For given a desired allocation $\bm\rho(\bm\theta)$, the asymptotic variability of the sample allocation proportions $\bm N_n/n$  is perhaps the most important issue.   The variability of sample allocation proportions can have a strong effect on power. This
has been demonstrated by many simulation studies of Melfi and Page (1998)
and Rosenberger et al.  (2001),  and  theoretically by Hu and Rosenberger (2003), who show explicitly the
relationship between the power of a test and the variability of the randomization
procedure for a given allocation proportion. In the latter paper, the authors showed that the
average power of a randomization procedure is a decreasing function of the
variability of the procedure.   Hu, Rosenberger and Zhang (2006) introduced {\em asymptotically best} response adaptive
randomization procedures as ones that have the smallest variance of the
allocation proportion among the procedures targeting the same allocation. These   results allow a formal
assessment of operating characteristics of various response-adaptive randomization
designs. The following theorem of Hu, Rosenberger and Zhang (2006) shows that for any given an allocation proportion, there is a lower bound of the asymptotic variability of the sample allocation proportions which converge to this given allocation proportion.

\begin{theorem} Assume the following regularity conditions:
\begin{enumerate}
  \item  The parameter space $\bm\Theta$ of $\bm\theta$ is an open set;
  \item  The distributions of responses $f_1(\cdot|\bm\theta_1),\ldots,f_K(\cdot|\bm\theta_1)$ follow an exponential family;
  \item  For the limiting allocation proportion $\bm\rho(\bm\theta)=\big(\rho_1(\bm\theta),\ldots,\rho_K(\bm\theta)\big)$,
  $$ \frac{N_{n,k}}{n}\to \rho_k(\bm\theta)\; a.s. \;\; k=1,\ldots, K; $$
  \item For a positive definite matrix $\bm V(\bm\theta)$,
  $$ \sqrt{n}\Big(\frac{N_n}{n}-\bm\rho(\bm\theta)\Big)\overset{D}\to N\big(\bm 0,\bm V(\bm\theta)\big). $$
\end{enumerate}
Then there exists a $\bm\Theta_0\subset \bm\Theta$ with Lebesgue measure $0$ such that for every $\bm\theta\in \bm\Theta\setminus \bm \Theta_0$,
$$ \bm V(\bm\theta)\ge \bm\Sigma_{LB},$$
where  $ \bm\Sigma_{LB}$  is defined by (\ref{eqLB}).
\end{theorem}
We refer to an adaptive design that attains the lower bound as {\em asymptotic best} (or {\em efficient}) for that particular allocation $\bm\rho(\bm\theta)$. Table \ref{tab3} gives the lower bounds of the asymptotic variabilities  $\sigma_{LB}^2$  for the urn proportion $\frac{q_2}{q_1+q_2}$, the RSIHR proportion  $\frac{\sqrt{p_1} }{\sqrt{p_1}+\sqrt{p_2}}$ and the Neyman proportion $\frac{\sqrt{p_1q_1}}{\sqrt{p_1q_1}+\sqrt{p_2q_2}}$ in a two-treatment clinical trial with binary responses.
\begin{table}[h]\begin{center} \renewcommand{\arraystretch}{1.3}
\vspace{0.2cm}
\begin{tabular}{|c|c|c|}
  \hline
  & $\rho_1(\bm\theta)$ & $\sigma^2_{LB}$ \\
   \hline
  Urn proportion  &   $\frac{q_2}{q_1+q_2}$ & $\frac{q_1q_2(p_1+p_2)}{(q_1+q_2)^3}$ \\
 RSIHR proportion &  $\frac{\sqrt{p_1} }{\sqrt{p_1}+\sqrt{p_2}}$ & $\frac{1}{4(\sqrt{p_1}+\sqrt{p_2})^3}\left(\frac{p_2q_1}{\sqrt{p_1}}+\frac{p_1q_2}{\sqrt{p_2}}\right)$ \\
 Neyman proportion &  $\frac{\sqrt{p_1q_1}}{\sqrt{p_1q_1}+\sqrt{p_2q_2}}$  &  $\frac{1}{4(\sqrt{p_1q_1}+\sqrt{p_2q_2})^3}\left(\frac{p_2q_2(1-2p_1)^2}{\sqrt{p_1q_1}}+\frac{p_1q_1(1-2p_2)^2}{\sqrt{p_2q_2}}\right)$  \\
  \hline
  \end{tabular}\end{center}
\caption{Lower bounds  of RAR procedures for several allocation proportions. }
\label{tab3}\end{table}

For the case of two-treatment trials with binary responses, among the randomization procedures as given in the last subsection which target the same allocation $q_2/(q_1+q_2)$, the RPW rule has the largest variability which is faraway from the lower bound, and the DL rule is an asymptotically best response adaptive
randomization procedure.
For a general allocation proportion $\bm\rho(\bm\theta)$, only the DBCD can approach an asymptotically best response adaptive
randomization procedure as $\gamma\to \infty$ (Table \ref{tab2}).

\subsection{Efficient RAR Designs}
  Hu, Zhang and He (2009) proposed a class of {\em efficient response adaptive randomization designs}, which are fully randomized, can target any allocation (under mild
regularity conditions), and are asymptotically best. The ERADE was proposed for comparing two treatments. After $m$ assignments, we let $\widehat{\bm \theta}_m=(\widehat{\bm\theta}_{m,1}, \widehat{\bm\theta}_{m,2})$ be the MLE of the parameter $\bm\theta=(\bm\theta_1,\bm\theta_2)$. The   probability of assigning the $(m+1)$-th patient to treatment $1$ is defined by
\begin{equation}\label{eqERADE1} p_{m+1,1}= \begin{cases} \alpha \rho_1\big(\widehat{\bm\theta}_m\big), & \text{ if } \frac{N_m}{m}> \rho_1\big(\widehat{\bm\theta}_m\big),\\
  \rho_1\big(\widehat{\bm\theta}_m\big), & \text{ if } \frac{N_m}{m}= \rho_1\big(\widehat{\bm\theta}_m\big),\\
 1-\alpha \big(1-\rho_1\big(\widehat{\bm\theta}_m\big)\big), & \text{ if } \frac{N_m}{m}< \rho_1\big(\widehat{\bm\theta}_m\big),
 \end{cases}
 \end{equation}
where $0\le \alpha<1$ is a pre-specified constant.  When $\rho(\bm\theta)\equiv 1/2$, the procedure reduces to  the famous  Efron's biased coin design (Efron (1971)).
The constant $\alpha$   is related to the randomness of the design. When $\alpha=1$, the procedure reduces to the SMLP.
When  $\alpha$ is smaller, the ERADE is more deterministic and could have smaller variability. Hu, Zhang and He (2009) recommend  choosing $\alpha$ between $0.4$ and $0.7$.
\begin{theorem} Suppose the distributions $f_1(\cdot|\bm\theta_1)$ and $f_2(\cdot|\bm\theta_2)$ of the responses follow an exponential family.  Then
$$  N_{n,1}-n\rho_1 =O(\sqrt{n\log\log n})\; a.s. \;\; \text{ and } \; $$
$$ \max_{m\le n} \left| N_{m,1}-m\rho_1(\widehat{\bm\theta}_m)\right|=o(\sqrt{n}) \; \text{ in probability}, $$
where $\rho_1=\rho_1(\bm\theta)$.
In particular,
$$ \sqrt{n}\Big(\widehat{\bm\theta}_n-\bm\theta, \frac{N_{n,1}}{n}-\rho_1\Big)
\overset{D}\to N\left(\bm 0,\bm\Lambda\right)\;\;\text{ with } $$
$$\bm\Lambda= \begin{pmatrix} \bm I^{-1}(\bm\theta) & \bm I^{-1}(\bm\theta)\frac{\partial \rho_1(\bm\theta)}{\partial \bm\theta} \\
\big( \frac{\partial \rho_1(\bm\theta)}{\partial \bm\theta}\big)^{\prime}  \bm I^{-1}(\bm\theta) &   \sigma^2_{LB}\end{pmatrix}, \;\;
\sigma^2_{LB}=\big( \frac{\partial \rho_1(\bm\theta)}{\partial \bm\theta}\big)^{\prime}  \bm I^{-1}(\bm\theta) \frac{\partial \rho_1(\bm\theta)}{\partial \bm\theta}.  $$
\end{theorem}

\bigskip

 Under the situation where efficiency is critically important, in theory, the ERADE should be the best choice in all the response-adaptive randomization procedures.   The simulation evidence that the ERADE outperforms other procedures in most cases  can be found in Hu, Zhang and He (2009), Flournoy, Haines and Rosenberger (2014) etc.  When the responses are binary and the desired proportion is $q_2/(q_1+q_2)$, the DL rule is also an efficient response-adaptive randomization procedure.    However, simulations shows that the DL rule
produces an allocation that is suboptimal with respect to power. Its finite-sample variances are much smaller than the
corresponding asymptotic variances, but when $p_1$ and $p_2$ are large and different,
the DL rule does not converge to the target allocation proportion as fast as
other procedures (c.f. Hu, Zhang and Hu (2009)). Similar simulation evidence is also found for the ERADE  under some situations  when the sample size is small. It sometimes does not converge to the target allocation proportion as fast as the DBCD does, though its finite-sample variances are always small.
The main reason for such a phenomenon   is possibly that the allocation probabilities in the DL rule and the the ERADE are not stable, they always jump from one value to another. A continuous allocation function can make the allocation probabilities stable and speed up the convergence of the sample allocation proportions.

 Very recently, Zhang et al. (2014b) proposed a new ERADE for multi-treatment trials by defining a continuous allocation function. Let $\psi(x)$ be a weight function given by
  $$\psi(x)=1+\sqrt{\big(x^{2\gamma}-1)\vee 0}, \;\; x\ge 0. $$
  Define the allocation function $\bm g(\bm x,\bm y)=\big( g_1(\bm x,\bm y),\ldots,  g_K(\bm x,\bm y)\big)$ by
\begin{equation}\label{eqERADE2}  g_k(\bm x,\bm y)=\frac{y_k\psi\big(\frac{y_k}{x_k}\big)}{\sum_{j=1}^K y_j\psi\big(\frac{y_j}{x_j}\big)}, \;\; k=1,\ldots, K.
\end{equation}
 After $m$ assignments, the probability of assigning the $(m+1)$-th patient to treatment $k$ is defined by
 $$ p_{m+1,k}=g_k\left(\frac{\bm N_m}{m}, \widehat{\bm \rho}_m\right)\; \text{ with } \widehat{\bm \rho}_m=\rho(\widehat{\bm\theta}_m),\;\; k=1,\ldots, K. $$
\begin{theorem} Suppose the distributions $f_1(\cdot|\bm\theta_1), \ldots, f_K(\cdot|\bm\theta_K)$ of the responses follow an exponential family.  Then $p_{m+1,k}\to \rho_k$, $k=1,\ldots, K$,
$$  \bm N_n-n\bm \rho=O(\sqrt{n\log\log n})\; a.s. \;\; \text{ and } \; $$
$$ \max_{m\le n} \left\| \bm N_m-m\bm \rho(\widehat{\bm\theta}_m)\right\|=o(\sqrt{n}) \; \text{ in probability}, $$
where $\bm\rho=(\rho_1,\ldots,\rho_K)=\bm\rho(\bm\theta)$. Further there is a multi-dimensional standard Brownian motion  $\bm W(t)$ such that
$$ n(\widehat{\bm \theta}_n-\bm\theta)=\bm W(n)\bm I^{-1/2}(\bm\theta)+o(\sqrt{n}) \;\; a.s. \;\text{ and } $$
$$ \max_{m\le n} \left\| \bm N_m-m\bm\rho -\bm W(m)\bm I^{-1/2}(\bm\theta)\frac{\partial \bm\rho(\bm\theta)}{\partial \bm\theta}\right\|=o(\sqrt{n}) \; \text{ in probability}. $$
In particular,
$$ \sqrt{n}\Big(\widehat{\bm\theta}_n-\bm\theta, \frac{\bm N_n}{n}-\bm\rho\Big)
\overset{D}\to N\left(\bm 0,\bm\Lambda\right)\;\;\text{ with } $$
$$\bm\Lambda= \begin{pmatrix} \bm I^{-1}(\bm\theta) & \bm I^{-1}(\bm\theta)\frac{\partial \bm\rho(\bm\theta)}{\partial \bm\theta} \\
\big( \frac{\partial \bm\rho(\bm\theta)}{\partial \bm\theta}\big)^{\prime}  \bm I(\bm\theta) &   \bm\Sigma_{LB}\end{pmatrix}, \;\;
\bm\Sigma_{LB}=\big( \frac{\partial \bm\rho(\bm\theta)}{\partial \bm\theta}\big)^{\prime}  \bm I^{-1}(\bm\theta) \frac{\partial \bm\rho(\bm\theta)}{\partial \bm\theta}. $$
\end{theorem}

\section{Selection Bias and  Lack of Randomness}\label{sectSelBias}
\setcounter{equation}{0}
If the experimenter can predict the next
assignment, he may consciously or unconsciously bias the
experiment as to what treatment particular types of subjects should receive. Randomization   is used for neutralizing such bias in clinical trials.
 A natural measure of the selection bias of a sequential design is  the
expected percentage of correct guesses the experimenter can make if he guesses optimally (c.f.,  Efron (1971)).
 Let $J_m =1$ if the $m$th assignment
is guessed correctly, and $J_m =0$ otherwise. The expected proportion of correct guesses is the
expected value of $ \frac{1}{n}\sum_{m=1}^n J_m. $
So  selection bias of the design is defined by
 $$ SB_n=\ep\left[\frac{1}{n}\sum_{m=1}^n J_m\right]=\frac{1}{n}\sum_{m=1}^n \pr(J_m=1). $$
It is obvious that, for a complete randomization, in which each patient is assigned to each one of the $K$ treatments with the same probability $1/K$, the  selection bias is $SB_n=1/K$ which is the smallest value of the selection biases  of all randomization procedure in a trail with $K$ treatments. The selection bias can be regarded as a measure of lack of randomness.

  In a response-adaptive randomization procedure, the optimal guessing strategy is to guess  treatment $k$ for which
  $ p_{m+1,k}=\max_j p_{m+1,j}. $
So
$$P(J_{m+1}=1)=\ep [\max_k p_{m+1,k}]. $$
It is obvious that $\frac{1}{K}\le SB_n\le 1$. When the allocation is not balance, the optimal value $1/K$ can not be attained.   Zhang et al. (2014b) gives the minimum value of the asymptotic selection bias.

\begin{theorem}\label{thSB} For any adaptive design, if $\bm N_n/n\to \bm \rho(\bm\Theta)$ in probability,
then
 $$\liminf_{n\to \infty} SB_n\ge \max_k \rho_k(\bm \Theta).$$
 Further, if $p_{m+1,k}\to \rho_k(\bm\theta)$ in probability, $k=1,\ldots, K$, then
 $$ SB=:\lim_{n\to \infty} SB_n= \max_k \rho_k(\bm \Theta). $$
 \end{theorem}
According to this theorem, $\max_k \rho_k(\bm \Theta)$ is the minimum value of the asymptotic selection bias with a target allocation proportion $\bm \rho(\bm\Theta)$.
And, if the allocation probabilities converge to the target allocation proportion, then the design attains the lower bound of the selection bias. As a conclusion, the GPU, the SEU, the DBCD, the SMLP (as a special case of DBCD) and the new ERADE of Zhang et al. (2014b) attains the lower bound of the selection bias. However, the asymptotic selection bias $SB$ of the Hu, Zhang and He (2009) procedure is a  monotone function of the parameter $\alpha$, which coincides the intuition via the definition of the allocation function.
\begin{theorem} \label{thSBHuZhHe} Consider the procedure of  Hu, Zhang and He (2009) for two-treatment clinical trials.    Suppose that  $\pr(\rho_1(\widehat{\bm\Theta}_m)=v)= 0$   for any rational $v\in[0,1]$.   Then  we have
$$SB=\left\{\begin{array}{ll}
  1-2\alpha\rho_1\rho_2,  &  \text{ if } \rho_1\vee \rho_2 \le \frac{1}{2\alpha}, \\
 \rho_1\vee \rho_2, &  \text{ if } \rho_1\vee \rho_2\ge \frac{1}{2\alpha}.
  \end{array}\right.$$
\end{theorem}
It turns out that only the new ERADE of Zhang et al. (2014b) achieves  both the lower bound of asymptotic variability and the lower bound of the selection bias.

Recall that for a sequence $Y_1,Y_2,\ldots, Y_n$ of random variables taking values $0$ and $1$ with mean $p$, they are completely  random if and only if
$\ep[Y_i|\mathscr{F}_{i-1}]=p$, $i=1,\ldots, n$. So
$$ \frac{1}{n}\sum_{i=1}^n \ep\big|\ep[Y_i|\mathscr{F}_{i-1}]-p\big|, $$
 can be regarded as a measure of the lack of randomness of this sequence.
Therefore, for a response-adaptive randomization procedure, we can define another measure of  the lack of randomness by
 $$ MLR_n= \frac{1}{n} \sum_{m=1}^n\frac{1}{K}\sum_{k=1}^K \ep\big|p_{m,k}-\rho_k|. $$
 It is obvious that  if $p_{m,k}\to \rho_k$ in probability, $k=1,\ldots,K$, then
 $$MLR=:\lim_{n\to \infty} MLR_n=0. $$
 So, for the GPU, the SEU, the DBCD and the new ERADE of Zhang et al. (2014b), $MLR=0$.
 For the procedure of Hu, Zhang and He (2009), $MLR=2(1-\alpha)\rho_1\rho_2$ is a also a decreasing function of $\alpha$. The conclusion  coincides with that for $SB$.

  For further considering the degree of the lack of randomness of the designs with $MLR=0$, we take the case of $K=2$ as example.
  Recall that in the RPW rule,
  $$ p_{m+1,1}=\frac{Y_{m,1}}{m}\overset{D}\sim N\big(\rho_1, \frac{\widetilde{\sigma}_{RPW}^2}{m}\big), \;\;\widetilde{\sigma}_{RPW}^2= \frac{q_1q_2}{(2(q_1+q_2)-1)(q_1+q_2)^2},$$
  when $q_1+q_2>1/2$, and $\widetilde{\sigma}_{RPW}^2=\infty$ when $q_1+q_2\le 1/2$.  So,
  $$\lim_{n\to \infty}\sqrt{n}(MLR_n)=\lim_{n\to \infty} \frac{1}{\sqrt{n}}\sum_{m=1}^n \frac{\sqrt{2/\pi}\widetilde{\sigma}_{RPW}}{m^{1/2}}=\sqrt{8/\pi}\,\widetilde{\sigma}_{RPW}, $$
  when $q_1+q_2>1/2$, and  $\lim_{n\to \infty}\sqrt{n}(MLR_n)=\infty$ when $q_1+q_2\le 1/2$.
  Similarly, for the DBCD,
  $$\lim_{n\to \infty}\sqrt{n}(MLR_n)=  \sqrt{8/\pi}\,\widetilde{\sigma}_{DBCD},\;\; \widetilde{\sigma}_{DBCD}^2=\frac{\gamma^2\rho_1\rho_2+(1+\gamma)^2\sigma^2_{LB}}{1+2\gamma}. $$
  Because  $\widetilde{\sigma}_{DBCD}^2$ is an  increasing function of $\gamma$,   the large is $\gamma$, the more is the lack of randomness of the design. When $\gamma=0$, the procedure is the most random  one in the family of DBCDs and $\lim_{n\to \infty}\sqrt{n}(MLR_n)$ takes the smallest value $\sqrt{8/\pi}\sigma_{LB}$.  This coincides the intuition via the definition of the allocation function. It is easily to show that in the binary response case with target allocation $q_2/(q_1+q_2)$, $\widetilde{\sigma}_{RPW}$ is also greater than $\sigma^2_{LB}$.  We conjecture that in the class of all adaptive randomization procedures with a same allocation target, the DBCD with $\gamma=0$ (i.e. the SMLP) is the most random one.
When $\gamma=0$, the allocation probability is $\rho_1(\widehat{\bm\theta}_m)$ which is an asymptotically best estimator of $\rho_1(\bm\theta)$, and so the adaptive randomization procedure is {\em locally efficient}.

  For the new ERADE of Zhang et al. (2014b), we have
$  \lim_{n\to \infty}\sqrt{n}(MLR_n)=\infty. $

  Hence, the efficiency and the randomness are always two conflicting  features of adaptive randomization designs. However, it is  surprising that  the variability of the RPW rule is much higher than other designs including the DBCD,  and at the same time,  the RPW rule is  much less random than the DBCD when $p_1+p_2$ is close to or exceeds $1.5$. The value of $SB$ and $MLR$ for the DL rule, the GDL rule and the IMU design is an open problem. It is very possibly a positive constant depending the initial value  of the
   urn composition.

  \section{Survival and Delayed Responses}\label{sectDelay}
\setcounter{equation}{0}
In the  framework of response-adaptive randomization designs discussed above, the outcomes of the treatments are assumed to be complete and available immediately. In practices, many clinical trials have time-to-event
outcomes and the outcomes may be censored. The problem of handling survival responses in response-adaptive randomization designs have been studied by Rosenberger and Seshaiyer (1997),  Zhang and Rosenberger (2007),  Sverdlov  Ryeznik and Wong (2014)  etc.  For applying the response-adaptive randomization methodology to survival trials, the first problem is how to derive the desired allocation under reasonable optimization criteria.  Several meaningful  optimization methods  have been proposed by Zhang and Rosenberger (2007) and Sverdlov  Ryeznik and Wong (2014) in the framework of parameter survival models. The other problem is that, when the outcomes are time-to-event,  they are usually delayed and will not be observed before the next step of the sequential  procedure is carried out.  The delay time is usually the observed time or the censoring time.  The estimation of the parameters  and
the updatie of the urn composition (when using urn models), can only be processed according to
observed responses. The effect of the delay of treatment results is fist studied in theory
by Bai, Hu and Rosenberger (2002) for the urn compositions in   GPU designs. After that, Hu and Zhang (2004b), Zhang et al.  (2007),
Sun, Cheung and Zhang (2007) and Hu et al. (2008) has showed that the delay machine does not effect the asymptotic properties
of the sample allocation proportions for many adaptive designs if the delay degree
decays with a power rate. The basic reason is that the total delayed responses is
a high order infinitesimal of square root of the sample size when the delay degree decays with a
power rate.

To describe the delay machine, we let $t_m$ be the entry time of the $m$-th patient,
where $t_m$ is an increasing sequence of random variables. Assume that $\{t_{m+1}- t_m\}$ is
a sequence of independent random variables. The response time of the $m$-th patient
on treatment $k$ is denoted by $r_m(k)$. Suppose $\{r_m(k);m\ge 1\}$ is a sequence of independent
random variables, $k=1,\ldots, K$.  Further, assume that $\{t_{m+1}-t_m, r_m(k); k=1,\ldots,K,m\ge n\}$  is independent of the assignments
$\bm X_1, \ldots, \bm X_n$.
\begin{assumption}\label{assumpDelay} Let $\delta_k(m,\ell)=I\{r_m(k)\ge t_{m+\ell}-t_m\}$ be an indicator function that takes the value $1$ if the outcome of the $m$-th patient on treatment $k$ occurs after at least another $\ell$ patients arrive at the trials, and $0$ otherwise. Suppose for some constans $C>0$ and $\beta\ge 2$,
$$ \mu_k(m,\ell)=\pr\big(\delta_k(m,\ell)=1\big)\le C \ell^{-\beta}, \text{ for all } m,\ell, k.$$
\end{assumption}
This assumption is widely satisfied. A practical approach is to assume that the entry
mechanism generates a Poisson process and the delay time has an exponential distribution
in which both $\{r_m(k)\}$ and $\{t_{m+1}-t_m\}$ are sequences of i.i.d. exponential
random variables with means $\lambda_k>0$ and $\lambda_0>0$, respectively. This approach is
common in clinical studies and the probability $\mu_k(m,\ell)$ is $\big(\lambda_k/(\lambda_0+\lambda_k)\big)^{\ell}$.

Let $S^{obs}_{m,k}$ (resp. $N^{obs}_{m,k}$)  be the summation (resp. the number) of the outcomes on treatment
$k$ observed prior to the $(m+1)$-th assignment, and $S_{m,k}$ (resp. $N_{m,k}$) be the summation  (resp.
the number) of all the outcomes of those being assigned to treatment $k$ in the first $m$ patients, $k=1,\ldots, K$. The total delayed responses are
then $S_{m,k}-S_{m,k}^{obs}$, $k=1,\ldots, K$.   Then we have the following theorem.

\begin{theorem} Suppose Assumption  \ref{assumpDelay} is satisfied, and the responses on each treatment
are i.i.d. random variables having finite $(2+\delta)$-th moments. Then for some $0<\delta_0<\frac{1}{2}-\frac{1}{2+\delta}$, we have
$$ S_{m,k}-S_{m,k}^{obs}=o(n^{1/2-\delta_0}) \;\; a.s. $$
and
$ N_{m,k}-N_{m,k}^{obs}=o(n^{1/2-\delta_0})$ a.s.
\end{theorem}

\section{Other RAR Designs and Concluding Remarks}\label{sectConclud}
\setcounter{equation}{0}
Besides the response-adaptive randomization designs disused in this paper, many other procedures have been proposed in literature. For example, Thompson (1933) described a Bayesian procedure to compute the probability that one treatment is better than another with binary responses. Under a uniform prior distribution, the procedure yields the following formula. Given sufficient statistics $(S_{m,1},S_{m,2},N_{m,1}, N_{m,2})$ after $m$ patients, where $S_{m,1}$, $S_{m,2}$ are the numbers of successes of treatment $1$ and $2$ respectively, then
$$ \widehat{P}(p_1>p_2)
=\frac{\sum_{a=0}^{S_{m,1}}\binom{S_{m,1}+S_{m,2}-a}{S_{m,2}}\binom{N_{m,1}-S_{m,1}+N_{m,2}-S_{m,2}}{N_{m,2}-S_{m,2}}}{\binom{n+2}{N_{m,2}+1}}. $$
Thall and Wathen (2007) reformulated this procedure by proposing the following response-adaptive procedure:
$$ p_{m+1,k}=\frac{ \big[\widehat{P}(p_1>p_2)\big]^c}{\big[\widehat{P}(p_1>p_2)\big]^c+\big[1-\widehat{P}(p_1>p_2)\big]^c}. $$
They suggested selecting $c=(N_{m,1}+N_{m,2})/(2n)$ and updating $c$ adaptively to reduce the root MSE of the procedure.

Various designs are usually proposed in different views. Note that a design is ``good'' under one criterion may be not ``good'' under another criterion. A graphical comparison of response-adaptive randomization procedures can be found in Flournoy, Haines and Roenberger (2014).

In this paper, we give
 an overview of important response-adaptive randomization designs and their asymptotic theory. In practice, a successful application of a response-adaptive randomization design in  a clinical trial   may depend on at least three  essentials.  The first one is a good adaptive procedure. We suggest that the procedure should be fully randomized, have small   variabilities and converges fast  to an allocation  which is optimal under a given optimization criterion. The main drawback of the RPW is that it has high variability. Also, its allocation proportion is not an optimal allocation under  opportune optimization criteria.
  The second essential is a good estimation method to collect the information from the data on time. The sufficient statics and MLE are usually suggested.
  When the sample size is small, the Bayesian method  is a reasonable candidate.  Other estimators such as weighted likelihood estimators, robust estimators can be also used. The third essential is a not-worse start of the procedure. The performance of the RPW and urn models depend heavily  on the initial   composition of the urn,  especially when the sample size is not large. In a pediatric trial of extracorporeal membrane
oxygenation (ECMO; Bartlett et al., 1985), the RPW rule with 2 initial balls, 1 of each type, was used to randomize the patients. After several initial assignments in which the balls added were all of the ECMO type, the urn composition became very extreme so that the conventional therapy has very little  chance to be selected.   If the  RPW rule with 10 initial balls, 5 of each type, had been  used, maybe the story would have been different. The convergence   of the urn proportions and the sample allocation proportions for the RPW  rule is very slow. The DBCD and the ERADE converge much faster than the urn models. But also  the initial estimator of the parameters may have some effects on their operating
characteristics, especially when the sample size is very small. A  relatively accurate initial estimator may improve the performance of the design. If there is not enough information to get a good estimator at the initial several steps, the estimator or the allocation probabilities  are suggested to be modified such that the randomization is close to the restricted randomization until a rwasonable quality estimator is available. For details of the discuss about the applications of response-adaptive randomization designs, one can refer to  Rosenberger, Sverdlov and Hu  (2012) and
Sverdlov and Rosenberger (2013b).

 \bigskip

\baselineskip 14pt \linewidth 6.8in \oddsidemargin=-0.1in
\begin{center}
REFERENCES
\end{center}
\baselineskip = 14pt

\begin{verse}
\vspace{-0.28in} \hspace{-0.5in} {\small \baselineskip 14pt

\item Bai, Z. D., Hu, F. (2005). Asymptotics in randomized urn models. {\em Annals of Applied
Probability}, {\bf 15}: 914-940.

\item Bai, Z. D., Hu, F., Rosenberger, W. F. (2002). Asymptotic properties of adaptive designs
for clinical trials with delayed response. {\em Annals of Statistics}, {\bf 30}: 122-139.

\item
 Bai, Z. D., Hu, F. and Zhang, L.-X.  (2002).
The Gaussian approximation theorems for urn models and their
applications. {\em Annals of Applied  Probability}, {\bf 12}: 1149-1173.

\item Baldi Antognini, A., Giovagnoli, A. (2010). Compound optimal allocation for individual and
collective ethics in binary clinical trials. {\em Biometrika}, {\bf 97}: 935-946.

\item Bartlett, R. H., Rolloff, D. W., Cornell, R. G., Andrews, A. F., Dillon, P.
W., Zwischenberger, J. B. (1985). Extracorporeal circulation in neonatal
repiratory failure: a prospective randomized trial. {\em Pediatrics}, {\bf 76}: 479-487.

\item Biswas, A., Bhattacharya, R. (2009). Optimal response-adaptive designs for normal
responses. {\em Biometrical Journal}, {\bf  51}: 193-202.

\item Biswas, A., Bhattacharya, R. (2010). An optimal response-adaptive design with dual
constraints. {\em Statistics and Probability Letters}, {\bf 80}: 177-185.

\item Biswas, A., Bhattacharya, R. (2011). Optimal response-adaptive allocation designs in phase
III clinical trials: Incorporating ethics in optimality. {\em Statistics and Probability Letters},
{\bf 81}: 1155-1160.

\item Biswas, A., Mandal (2004). Optimal adaptive designs in pase III clinical trials for continuuos
responses with covariates. In: Bucchianico, A. Di., Lauter, H., Wynn, H. P., eds.
{\em mODa 7-Advances in Model-Oriented Design and Analysis}. Heidelberg: Physica-Verlag,
pp. 51-58.

\item Eisele, J. R. (1994). The doubly-adaptive biased coin design for sequential clinical trials.
{\em Journal of Statistical Planning and Inference}, {\bf 38}: 249-262.

\item Efron, B. (1971). Forcing a sequential experiment to be balanced. {\em Biometrika}, {\bf 58}: 403-417.

\item Gwise, T., Hu, J., Hu, F. (2008). Optimal biased coins for two-arm clinical trials. {\em Statistics
and its Interface}, {\bf  1}: 125-135.

\item Flournoy, N., Haines, L. M., Rosenberger, W. F. (2013). A graphical
comparison of response-adaptive randomization procedures. {\em Statistics in
Biopharmaceutical Research}, {\bf 5}: 126-141.

\item Gwise, T., Zhu, J., Hu, F. (2011). An optimal response adaptive biased coin design
with $k$ heteroscedastic treatments. {\em Journal of Statistical Planning and Inference}, {\bf  141}: 235-242.

\item Hu, F., Rosenberger, W. F. (2003). Optimality, variability, power: Evaluating responseadaptive
randomization procedures for treatment comparisons. {\em Journal of the American
Statistical Association}, {\bf  98}: 671-678.

\item Hu, F., Rosenberger, W. F. (2006). {\em The Theory of Response-Adaptive Randomization in Clinical
Trials}. New York: Wiley.

\item Hu, F., Rosenberger, W. F., Zhang, L.-X. (2006). Asymptotically best response- adaptive
randomization procedures. {\em Journal of Statistical Planning and Inference}, {\bf 136}: 1911-1922.

\item Hu, F., Zhang, L.-X. (2004a). Asymptotic properties of doubly-adaptive biased coin designs
for multitreatment clinical trials. {\em Annals of Statistics}, {\bf 32}: 268-301.

\item Hu, F., Zhang, L.-X. (2004b). Asymptotic normality of adaptive designs with delayed
response.{\em  Bernoulli},{\bf  10}: 447-463.

\item Hu, F., Zhang, L.-X., Cheung, S. H., Chan, W. S. (2008). Doubly-adaptive biased coin
designs with delayed responses. {\em Canadian Journal of Statistics}, {\bf  36}: 541-559.

\item Hu, F., Zhang, L.-X., He, X. (2009). Efficient randomized-adaptive designs. {\em Annals of
Statistics}, {\bf  37}: 2543-2560.

\item Ivanova, A. V. (2003). A play-the-winner type urn model with reduced variability. {\em Metrika},
{\bf 58}: 1-14.

\item Ivanova, A. V. (2006). Urn designs with immigration: Useful connection with continuous time stochastic
processes. {\em Journal of Statistical Planning and Inference}, {\bf 136}: 1836-1844.

\item Ivanova, A. V., Rosenberger, W. F. (2001). Adaptive designs for clinical trials with highly
successful treatments. {\em Drug Information Journal}, {\bf 35}: 1087-1093.

\item Jennison, C., Turnbull, B. W. (2000). {\em Group Sequential Methods with Applications to Clinical
Trials}. Boca Raton, FL: Chapman and Hall/CRC.

\item Jeon, Y., Hu, F. (2010). Optimal adaptive designs for binary response trials
with three treatments. {\em Statistics in Biopharmaceutical Research},  {\bf 2}(3):
310-318.

\item Li, W., Durham, S. D. and Flournoy, N. (1996). Randomized P\'olya urn designs. {\em Proceedings
of the Biometric Section of the Statistical Association}: 166-170.

\item May, C. and Flournoy, N. (2009). Asymptotics in response-adaptive designs generated by a
two-color, randomly reinforced urn. {\em Annals of  Statistics}, {\bf 37}(2): 1058-1078.

\item Melfi, V. and Page, C. (1998). Variability in adaptive designs for estimation of success probabilities.
In {\em  New Developments and Applications in Experimental Design} (N. Flournoy,
W. F. Rosenberger and W. K. Wong, eds.) 106-114. IMS, Hayward, CA.

\item Melfi, V. and Page, C. (2000). Estimation after adaptive allocation.  {\em Journal of Statistical Planning and Inference}, {\bf 87}:
353-363.

\item  Melfi, V. and Page, C. and Geraldes, M. (2001). An adaptive randomized design with application
to estimation. {\em Canadian Journal of Statistics}, {\bf  29}: 107-116.

\item Robbins, H. (1952). Some aspects of the sequential design of experiments. {\em Bulletin of the
American Mathematical Society}, {\bf 58}: 527-535.

\item Rosenberger, W. F. (2002). Randomized urn models and sequential design. {\em Sequential
Analysis}, {\bf  21}:1-41 (with discussion).

\item Rosenberger, W. F., Hu, F. (2004). Maximizing power and minimizing treatment failures in
clinical trials. {\em Clinical Trials}, {\bf 1}: 141-147.

\item Rosenberger, W. F., Lachin, J. L. (2002). {\em  Randomization in Clinical Trials: Theory and
Practice}. New York: Wiley.

\item Rosenberger, W. F., Seshaiyer, P. (1997). Adaptive survival trials. {\em Journal of
Biopharmaceutical Statistics},  {\bf 7}: 617-624.

\item Rosenberger, W. F., Stallard, N., Ivanova, A., Harper, C. N., Ricks, M.
L. (2001). Optimal adaptive designs for binary response trials, {\em Biometrics},
{\bf 57}, 909-913.

\item Rosenberger, W. F., Sverdlov, O., Hu, F. (2012). Adaptive randomization
for clinical trials. {\em Journal of Biopharmaceutical Statistics}, {\bf 22}: 719-736.

\item Sun, R., Cheung, S. H., Zhang, L.-X. (2007). A generalized drop-the-loser rule
for multi-treatment clinical trials. {\em Journal of Statistical Planning and Inference},
{\bf 137}: 2011-2023.

\item Sverdlov, O., Rosenberger, W. F. (2013a). On recent advances in optimal
allocation designs in clinical trials. {\em Journal of Statistical Theory and
Practice},  {\bf 7}, 753-773.

\item  Sverdlov, O., Rosenberger, W. F. (2013b). Randomization in clinical
trials: can we eliminate bias? {\em Clinical Investigation}, {\bf 3}(1), 37-47.

\item Sverdlov, O., Ryeznik, Y., Wong, W. K. (2012). Doubly adaptive biased
coin designs for balancing competing objectives in time-to-event trials.
{\em Statistics and Its Interface}, {\bf 5}: 401-413.

\item  Sverdlov, O., Ryeznik, Y., Wong, W. K. (2014). Efficient and ethical
response-adaptive randomization designs for multi-arm clinical trials with
Weibull time-to-event outcomes. {\em Journal of Biopharmaceutical Statistics},
{\bf 24}(4):732-754.

\item Sverdlov, O., Tymofyeyev, Y., Wong, W. K. (2011). Optimal response-adaptive randomized
designs for multi-armed survival trials. {\em Statistics in Medicine}, {\bf 30}: 2890-2910.

\item Thall, P. F., Wathen, J. K. (2007). Practical Bayesian adaptive randomization
in clinical trials. {\em European Journal of Cancer}, {\bf 43}: 860-867.

\item Thompson, W. R. (1933). On the likelihood that one unknown probability exceeds another
in the view of the evidence of the two samples. {\em  Biometrika},  {\bf 25}: 275-294.

\item Tymofyeyev, Y., Rosenberger, W. F., Hu, F. (2007). Implementing optimal allocation in
sequential binary response experiments. {\em Journal of the American Statistical Association},
{\bf 102}: 224-234.

\item Wei, L. J. (1979). The generalized P\'olya's urn design for sequential medical trials.{\em  Annals of Statistics}, {\bf 7}:
291-296.

\item Wei, L. J., Durham, S. (1978). The randomized play-the-winner rule in
medical trials. {\em Journal of the American Statistical Association}, {\bf 73}: 840-843.

\item Wong, W. K., Zhu, W. (2008). Optimum treatment allocation rules under
a variance heterogeneity model. {\em Statistics in Medicine}, {\bf 27}: 4581-4595.

\item Zelen, M. (1969). Play the winner rule and the controlled clinical trial.
{\em Journal of the American Statistical Association}, {\bf 64}: 131-146.

\item Zhang, L., Rosenberger, W. F. (2006). Response-adaptive randomization
for clinical trials with continuous outcomes. {\em Biometrics}, {\bf 62}(2):562-569.

\item Zhang, L., Rosenberger, W. F. (2007). Response-adaptive randomization
for survival trials: the parametric approach. {\em Applied Statistics}, {\bf 56}(2):
153-165.

\item Zhang, L.-X. (2012).  The Gaussian approximation for generalized Friedman's urn model with heterogeneous and unbalanced updating.
  {\em Science China-Mathematics}, {\bf 55}(11): 2379-2404.

\item Zhang, L.-X. (2014).  Central limit theorems for a recursive stochastic algorithm with applications to adaptive designs.
{\em Manuscript}.

\item Zhang, L.-X., Chan, W. S., Cheung, S. H., Hu, F. (2007). A generalized
drop-the-loser urn for clinical trials with delayed responses.{\em  Statistica
Sinica}, {\bf 17}: 387-409.

\item Zhang, L.-X. and Hu, F. (2009).
 The Gaussian approximation for multi-color generalized Friedman's urn model. {\em Science in China, Series A}, {\bf 52}(6): 1305-1326.

\item Zhang, L.-X., Hu, F., Cheung, S. H. (2006). Asymptotic theorems of
sequential estimation-adjusted urn models. {\em Annals of Applied Probability},
{\bf 16}:340-369.

\item Zhang, L.-X., Hu, F., Cheung, S. H., Chan, W. S. (2011). Immigrated
urn models|theoretical properties and applications. {\em The Annals
of Statistics}, {\bf 39}(1): 643-671.

\item Zhang, L.-X., Hu, F., Cheung, S.H. and Chan, W.S. (2014a).
 Asymptotic properties of multi-color randomly reinforced P\'olya urns. {\em Advances in Applied Probability}, {\bf 46}: 585-602.

\item Zhang, L.-X., Hu, F., Cheung, S. H., Chan, W. S. (2014b). Multiple-treatment efficient randomized-adaptive design with minimum selection bias.
{\em Manuscript}.

\item  Zhu, H., Hu F. (2009). Implementing optimal allocation for sequential
continuous responses with multiple treatments. {\em Journal of Statistical
Planning and Inference}, {\bf 139}: 2420-2430.

}
\end{verse}

 \end{document}